\documentclass[%
 reprint,
 amsmath,amssymb,
 aps,
]{revtex4-1}

\usepackage[utf8]{inputenc}
\usepackage{graphicx}
\usepackage{sidecap}
\usepackage{bigints}
\usepackage{amsthm}
\usepackage{framed}
\usepackage[font={footnotesize,it}]{caption}
\usepackage{color}
\usepackage{amssymb}
\theoremstyle{definition}

\newcommand{\rond}[2]{\, \frac{\partial #1}{\partial #2}}
\newcommand{\rondd}[2]{\, \frac{\partial^2 #1}{\partial #2^2}}
\newcommand{\rondx}[3]{\, \frac{\partial^2 #1}{\partial #2 \partial #3}}

\newcommand{\boldd}{\,\textbf}
\newcommand{\eq}[1]{\,\begin{equation}
                   #1 
                   \end{equation}
}

\newcommand{\morabba}[1]{\,\begin{flushright}
 \Rectsteel \\
\end{flushright}}

\newcommand{\eqq}[2]{\,\begin{equation} \label{#2}
                   #1 
                   \end{equation}
}

\newcommand{\al}[1]{\,\begin{align}
                   #1 
                   \end{align}
}
\newcommand{\all}[2]{\,\begin{align}
                   #1 
                    \label{#2}
                   \end{align}
}
\usepackage{setspace}

\usepackage[top=1.9cm, bottom=3.7cm, left=1.9cm, right=1.32cm]{geometry}

\begin{document}
\preprint{APS/123-QED}
\title{\boldd{Dynamics of Influence on Hierarchical Structures
}}

\author{Babak Fotouhi and Michael G. Rabbat \\
Department of Electrical and Computer Engineering\\
McGill University, Montr\'eal, Qu\'ebec, Canada\\
Email:\texttt{ babak.fotouhi@mail.mcgill.ca, michael.rabbat@mcgill.ca}\\}

\begin{abstract}
Dichotomous spin dynamics on a pyramidal hierarchical structure (the Bethe lattice) are studied.  The system embodies a number of  \emph{classes}, where a class comprises of nodes that are equidistant from the root (head node). Weighted links exist between nodes from the same and different classes. 
The spin (hereafter, \emph{state})  of the head node is fixed. We solve for the dynamics of the system for different boundary conditions. We find necessary conditions  so that the classes eventually repudiate or acquiesce in the state imposed by the head node.  The results indicate that to reach unanimity across the hierarchy, it suffices
that  the bottom-most class adopts the same state as the head node. Then the rest of the hierarchy will inevitably comply. This also sheds light on the importance of mass media as a means of synchronization between the top-most and bottom-most classes. Surprisingly, in the case of discord between the head node and the bottom-most classes, the average state over all nodes inclines towards that of the bottom-most class regardless of the link weights and intra-class configurations. Hence the role of the bottom-most class is signified.

\end{abstract}
\maketitle

\section{Introduction}

This paper discusses a generalization of the voter model~\cite{liggett} on a hierarchically-structured network. The underlying graph is a Bethe lattice with coordination number~$q$ (i.e., a complete $q$-ary tree), and we also allow for random connections between nodes at the same depth. The central focus of this paper is to identify the effect of the states adopted by nodes at the extreme ends of the hierarchy (the root, and the leaves in a finite network) on the dynamics and steady-state opinion of the system.

In the voter model, each node $x$ has a time-varying state $s_x(t)$ taking one of the values $\pm 1$. The states of nodes are considered to be binary to model, e.g., cases where individuals take a dichotomous stance on issues such as elections with two major parties. Nodes update their state in a randomized manner based on the states of their neighbors in the graph. Let $N_x$ denote the set of neighbors of node $x$ and let $J\{x' \rightarrow x\} \ge 0$ denote the time-homogeneous influence which a neighbor $x'$ has on $x$'s decision. Then the probability that $x$ flips its state at time $t$ is given by
\begin{equation}
w_x(t) = \frac{\sum_{x' \in N_x \colon s_{x'} \ne s_x} J\{x' \rightarrow x\} s_{x'}(t)}{\sum_{x' \in N_x} J\{x' \rightarrow x\}}.
\end{equation}

\begin{figure}
  \centering
  \includegraphics[width=1\columnwidth]{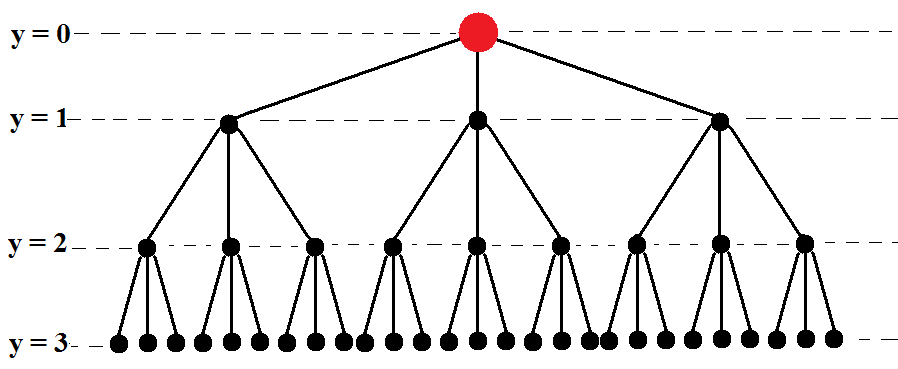}
  \caption[Figure ]%
  {\small {The first four levels of a hierarchy with $q=3$. The links whithin each class are not shown in the figure. The number of these links are modelled in two different ways in 
the sections below. First it is assumed to be a constant number, then a constant fraction of the possible links.} }
\label{tree}
\end{figure}

This paper focuses on hierarchically-structured networks. Specifically, we assume the network takes the form of a complete $q$-ary tree. We index depth in the tree using the variable $y$. There is a single node (the ``root'' or ``head'' node) at level $y=0$. The root is connected to $q$ nodes at level $1$; each of these nodes is connected to $q$ nodes at level 2, and so on (see Figure~\ref{tree}). Thus, there are $q^y$ nodes at level $y$, each with one connection to a parent at level $y-1$ and $q$ connections to children at level $y+1$. In addition, we allow for a node at level $y$ to have random connections to other nodes at the same level.

It is well known that for voter dynamics on general connected networks, when all nodes follow the dynamics described above, eventually a consensus is reached across the network on a single state, and the expected value of the consensus state depends on the initial conditions (fraction of the population initially with opinion $+1$ vs.~$-1$). If a single node is stubborn and fixes its state at the value $+1$ for all time, then eventually all nodes converge to the state $+1$. In general, when more than one node is stubborn and their opinions do not coincide, the opinions of the remaining nodes do not converge to a consensus; rather, they reach a steady-state where, in expectation, a fraction of the network holds $+1$, and size of this fraction depends on the opinions and locations of the stubborn agents within the network.

\subsection*{Contributions and Organization of the Paper}

We focus on the case where the root node fixes its opinion at $s_0(t) = +1$ for all time, and we investigate the effect this has on the steady-state opinions of the rest of the network under a mean-field approximation. For a node at level $y$, let the weights of influence, relative to weight $J\{y \rightarrow y\}$ that this node assigns to other nodes at the same level, be given by
\begin{equation}
\begin{cases}
\frac{J\{y-1 \rightarrow y\}}{J\{y \rightarrow y\}} = \beta \\
\frac{J\{y+1 \rightarrow y\}}{J\{y \rightarrow y\}} = \alpha.
\end{cases}
\end{equation}
That is, each parent exerts a relative influence of $\beta$ on its children, and children exert a relative weight of $\alpha$ on their parents. 

When only the root node is stubborn and holds its opinion fixed at $s_0(t) = +1$, then in the limit all nodes consent on the opinion $+1$ regardless of whether the network is finite (i.e., fixed depth $L < \infty$) or infinite (i.e., $L \rightarrow \infty$). We derive level-specific expressions for the rate of convergence in this setting. 

Then we consider the case where nodes at the bottom-most level of a finite network are also stubborn. Surprisingly, we find that when all of the nodes at the deepest level from the root have their states fixed at $s_L(t) = -1$, opposing the root node, then so long as $\beta > 0$, the majority of nodes in the network will have have negative state when $q > 1$, and when $q=1$ (i.e., the network is a chain of length $L$), then the majority opinion is the same as the sign of $\beta - \alpha$. Our findings highlight the importance of the role of the bottom-most class.

The rest of the paper is organized as follows. Section~\ref{sec:background} reviews background and provides motivation for the modeling assumptions adopted. Then two different models are considered. Section~\ref{sec:constantSelfWeight} examines a model where the self weights $J\{y \rightarrow y\}$ are all constant and independent of the level $y$. Section~\ref{sec:constantFraction} then considers the case where $J\{y \rightarrow y\}$ scales with the level so that the influence is proportional to the number of other nodes at level $y$. In both of these sections our analysis is based on a mean-field approximation where we allow both time $t$ and the level $y$ to be continuous valued. In Section~\ref{sec:discrete} we provide the analysis for the case where the level $y$ takes only integer values. We conclude in Section~\ref{sec:summary}.

\section{Background and Motivation} \label{sec:background}

\subsection{Socio-physics: The Ising and Voter Models }
Recently, statistical mechanics has offered its perspective of the micro-macro entwinement in problems originating from the social sciences. Many sociological problems  have been studied by physicists, through the vantage point of the Ising model~\cite{new1,  galam, lewenstein, sznajd, sznajd2, bernardes, contucci, bornoldt, galam2, stauffer, barra} and its dynamic generalizations~\cite{schweitzer2, krapivsky, vazquez2, slanina,martins, delre,naim1, masuda, lambiotte, vilone}. For a thorough review on these models, see~\cite{castel_rev, galam_rev, galambook}. For other instances of opinion dynamics and voting on hierarchical structures, see~\cite{Zgalam1, Zgalam2, Zgalam3, Zgalam4, galambook}. 

A kinetic generalization of the Ising model is  the voter model~\cite{liggett}.  
The rationale behind the conventional voter models is peer pressure and the influence of others on each agent, which   
 is evident in social interactions~\cite{sherif, asch,  milgram, zimbardo, kelman, cialdini, steve}.  

The conventional voter model considers a simple scenario in which,  
at each timestep, each node finds the fractions of its neighbors who currently agree and oppose with it, and with those probabilities follows either of the states. So if, for instance, all neighboring nodes are at state $+1$, 
the node definitely follows. However,  if $90 \%$ of them are at state +1, there is a $10\%$ chance that the node will adopt $-1$. 
Note that this model is easily generalized to allow for weights (a degree of influence) for each neighbor's opinion, as considered in this work.


The steady-state behavior of the voter model on arbitrary graphs is characterized in~\cite{sood} through a mean-field approximation. In~\cite{vazquez, volovik} the generalization to three states is considered. The $q$-voter model is introduced in~\cite{castellano}, where each node adopts the state of $q$ randomly selected neighbors, given that they are unanimous. The voter model on weighted graphs has been studied in~\cite{weighted1}; in that work a mean-field approximation is adopted whereby all nodes with the same degree hold the same state, and this is inapplicable to the problem addressed in this paper since the states depend on class level. 
See~\cite{caccioli, contra, castel_gen, rec1, rec2, rec3, rec4, rec5, rec6} for examples of recent generalizations of the voter model.


\subsection{Hierarchical Structures in Social Systems}

We next provide examples of cases where hierarchical systems appear in the social sciences to motivate the structural assumptions made in this paper. The first example is the hierarchical structure of organizations, the second one is the media industry, and the third one is the hierarchical nature of structures of power.

Modernization hinges on extensive rationalization of organizational and institutional structures. The first analytical treatment of modern bureaucratic organizations dates back to M.~Weber's seminal work~\cite{weber1}, in which he laid the foundation of the ideal type of modern bureaucracies and delineated how the properties of modern bureaucracies  increase efficiency. Among those traits of instrumental rationality 
 is the hierarchy of authorities,  which is central to this paper. Agents occupy different positions and have different degrees of power and influence over the organization. Bureaucratization of social institutions does increase efficiency in some ways, but comes with costs and drawbacks  (see for example~\cite{merton, blau1, blau2, peter, ouchi} for broader discussions). One of these pitfalls pertains to a problem examined in this paper called ``the iron law of oligarchy", a term coined by the sociologist R.~Michels~\cite{michels}. 

Oligarchy is the control of organizations of the society by the people at the summit (a.k.a., \emph{the elite}). It is a seemingly inevitable property of any large-scale bureaucratized organization. Michels contends that these organizations will monopolize power in the hands of those at the top. He concisely writes, ``It is organization which gives birth to the domination of the elected over the electors$\ldots$  Who says organization says oligarchy"~\cite{michels}. In this paper, we investigate the extent to which the state imposed by the head node will affect the state of other nodes at lower levels of the hierarchy.

The second  example that motivates studying diffusion of influence over hierarchical graphs is the network of mass media  organizations. Giant corporations own (or partially control) large media conglomerates which, in turn, control large TV networks or film studios and production companies,  which have subsidiary units of their own, and so on~\cite{greshon, shah, noam}. Hierarchy confines behavior; seldom is it the case that a low-ranked member of this hierarchy 
 reflects opinions or disseminates information  without approval (or at least the influence) of higher authorities, who in turn are controlled by the members at even higher levels. Hence it seems natural to assume that the agenda set by the ones at the top significantly  influences the action of the ones below~\cite{chomsky, bagdakian, mills1, dye,elmer}. For a thorough discussion on the effects that this type of hierarchical structure might have on the impartiality of the news and opinions reflected through the media, see~\cite{chomsky, bagdakian, anderson}. In particular, to get a sense of the extent to which hierarchical structures pervade  modern media networks, the reader is referred to~\cite[Chapter 1]{chomsky} and~\cite[Chapter 2]{bagdakian}.  

%
%
%

The third example in which hierarchical structures are evoked lies in the field political sociology. In his study on power structures and the power elite~\cite{mills1}, C.~Wright Mills perceived a pyramidal structure of power. He contended that there is a ruling class (the power elite) which is in a cohesively organized minority and controls the majority of the society in various economic, political and military realms. This motivates our use of a hierarchical model for society, with elites at the top and the masses at the bottom, and intermediary classes in the middle. As T.R. Dye argues~\cite{dye}, policy-making is a product of elite consensus, which is then imposed and, in turn, transformed into public demand, not the other way around. 


\subsection{The  Mean-Field Assumption } \label{sec:meanField}

In this section we motivate the  ``mean-field" assumption that will be adopted below. It falls in the domain of Marxian theory of class conflict~\cite{calhoun, appelrouth, coser}. Here we briefly describe its central notions.

Marxian social classification is based on capital and property ownership~\footnote{In this paper, it is tacitly assumed that there is a unique classifier of the members of society, based on which the levels are assigned. So the analysis is more applicable to situations where there is a dominant classification factor. For example, in an ideal-typical chain of command, one's rank is the determinant of influence, regardless of other socio-economic factors. 
}. 
This segregates the system into haves and have-nots, or the bourgeoisie 
and the proletariat (and a minority in the middle, i.e., the petit bourgeoisie).
Members of the same class share economic status and interests, which leads to common belief of affiliation, and consequently, convergence of action. Economic interests of the classes are of opposing natures. As disparity between the life conditions of the classes increases, class consciousness shifts into conflict and antagonism between the classes, and convergence of action is intensified. 
 This convergence of behavior is the rationale behind the assumption made in this paper, that nodes from the same class have the same state. 

 Class struggle further intensifies the pressure impinged upon the proletariat, which is then followed by a transient state called ``dictatorship of the proletariat", eventuating in a revolution. Hence the role of the bottom-most class of the society  is essential~\footnote{Let us mention that Weber's social classification adds other factors to that of Marx, such as status groups, yielding a classification with at least two dimensions, altering the consequences~\cite{weber7, giddens}.}. This paper also investigates the role of the bottom-most class and studies the effect they have on the collective state of the system. 

\section{First Model: Constant Self-weight
} \label{sec:constantSelfWeight}

Recall that $s_x(t) \in \{ \pm 1\}$ denotes the state of node $x$ at time $t$. Throughout this paper we adopt a mean-field assumption whereby the state of a node is replaced with the average state of all nodes at the same level. This assumption is structurally justified by observing the symmetric structure of the underlying network. Sociologically, this is equivalent to the behavioral convergence of class members mentioned in Section~\ref{sec:meanField} above. Based on this assumption, the tree collapses to a one-dimensional chain with one node representing each class. Let $s_y(t) \in [-1, +1]$ denote the average state of a node at level $y$ at time $t$. 

Recall also the strengths of influence that nodes at levels $y-1$ and $y+1$ have on nodes at level $y$,
\begin{equation}
\begin{cases}
\frac{J\{y-1\rightarrow y\}}{J\{y\rightarrow y\}}=\beta \\ \\
\frac{J\{y+1\rightarrow y\}}{J\{y\rightarrow y\}}=\alpha.
\end{cases}
\end{equation}
The model considered in this section supposes that each node also accounts for its own opinion with weight $p$. One can interpret $p$ as the number of neighbors a node has at the same level as itself (where neighbors are drawn uniformly from all nodes at the same level, to preserve symmetry), with the weight assigned to each of these neighbors being $J\{y \rightarrow y\} = 1$. Note that $p$ can be smaller than unity, which means that the weight it gives to members of the same level is less than those weights given  to members of neighboring levels.

The flipping probability of each node is proportional to the weighted fraction of its neighbors with the opposite state. 
Denote the flipping probability for nodes at level $y$ by $w_y(t)$. This means that with probability $w_y(t)$, we have $s_y(t+\Delta t)=-s_y(t)$, and with probability $1-w_y(t)$, we have $s_y(t+\Delta t)=s_y(t)$. Denote the set of nodes adjacent to the node at level $y$ by $N_y$. We have: 
\eq{
w_y= \displaystyle \frac{\displaystyle \sum_{x \in N_y, s_x \neq s_y} J\{ x\rightarrow y\} s_x}{\sum_{x \in N_y} J\{ x\rightarrow y\}}
,}
which simplifies to: 
\eq{ 
w_y=\displaystyle \frac{1}{2} \left[ 1- \displaystyle  \frac{s_y}{q\alpha + 1\times \beta + p  \times 1 } (\beta s_{y-1} + q \alpha s_{y+1} + p s_y ) \right]
.}
Using the flipping probability, we find the time evolution of $s_y(t)$:
\all{
\rond{{s}_y}{t}=  &
- \left( \frac{q\alpha+\beta}{q\alpha+\beta+p}\right) s_y \nonumber \\
& +\left( \frac{\beta}{q\alpha+\beta+p}\right) s_{y-1} + \left( \frac{q\alpha}{q\alpha+\beta+p}\right) s_{y+1} 
.
}{sy_1}

\subsection{Continuous Approximation, Semi-infinite Structure} 
Consider the case where the structure is limited from above but unlimited from below, i.e., $y\in  \mathbb{Z}^+=\{0, 1, 2, \ldots\}$. Note that~\eqref{sy_1} can be written equivalently as
\al{
\rond{s_y}{t}= & \left( \frac{q\alpha}{q\alpha+\beta+p}\right) (s_{y+1}+s_{y-1}-2s_y) \nonumber \\
&+\left( \frac{q\alpha -\beta}{q\alpha+\beta+p}\right) (s_y-s_{y-1})  
.
}
A  continuous approximation of (\ref{sy_1}) can be obtained as a differential equation in $y$ by relaxing $y$ to take values in $\mathbb{R^+}=\{x \in \mathbb{R}: x \geq 0 \}$, in which case we have
\eqq{ 
\partial_t s_y(t) = 
\left( \frac{q\alpha}{q\alpha+\beta+p}\right) \partial^2_{yy} s_y(t) +\left( \frac{q\alpha -\beta}{q\alpha+\beta+p}\right) \partial_y s_y(t) 
.}{sdot1}
This turns into the standard heat equation~\cite{polyanin} for $\phi(y,t)$  using the substitution: 
\eq{
s_y(t)=\phi(y,t) \exp\left[-\frac{(q\alpha-\beta)^2}{4 q \alpha (q\alpha+\beta+p)}t-\frac{(q\alpha-\beta)}{2q\alpha}y\right]
.}
If the initial state of the nodes are $s_y(0)$ and the head node is at some arbitrary state $s_{0}(t)$, defining 
$A \equiv q \alpha + \beta + p$ for brevity, then this equation has solution
\eqq{
s_y(t)=
\left( \sqrt{\frac{A}{q \alpha}} \right)\frac{\exp\left[-\frac{(q\alpha-\beta)^2 t}{4 q \alpha (A)}\right]}{2\sqrt{\pi t}}  \Psi(y,t) + \Phi(y,t)
,}{sy1}
where we have defined
\al{ \displaystyle
\begin{cases}
 \displaystyle \Psi(y,t)  &\stackrel{\text{def}}{=}
\bigintss_0^{\infty} s_{\xi}(t=0) \displaystyle \exp\left[\left( \frac{\xi-y}{2}\right) (\alpha q - \beta)\right] \nonumber \\ \\
\displaystyle &\! \! \!  \! \! \! \! \! \! \! \! \times \left\{ \displaystyle \exp\left[-\frac{(y-\xi)^2}{4 t q \alpha}(A)\right]-\displaystyle \exp\left[-\frac{(y+\xi)^2}{4 t q \alpha}(A)\right]   \right\} d \xi \\ \\ \\
\displaystyle \Phi(y,t) \! \! \!  &\stackrel{\text{def}}{=} \displaystyle \left( \frac{y}{2 \sqrt{\pi}} \sqrt{\frac{A}{q \alpha}} \right) \displaystyle \exp\left[\frac{\beta-q \alpha}{2q\alpha}y\right] \nonumber \\ \\
\displaystyle \! \! \! & \bigintss_0^t \displaystyle \frac{s_0(t-\lambda)}{\lambda^{\frac{3}{2}}}
 \displaystyle \exp\left[-\frac{(q\alpha-\beta)^2(\lambda)}{4 q \alpha(A) } - \frac{y^2(A)}{4 q \alpha \lambda}\right] d\lambda .
\end{cases}
}
The first term in~\eqref{sy1} is transient and is due to initial conditions. If we define 
\eq{
\gamma \stackrel{\text{def}}{=} \frac{|q\alpha-\beta|}{q \alpha + \beta + p}
,}
then $\Phi(y,t)$ can be approximated asymptotically for long times:
\al{
s_y(t) \sim \frac{\exp \left[ -\left( \frac{q\alpha-\beta}{2q\alpha}\right)y\right]}{\sqrt{\gamma y}}
 s_0\left(t- \gamma y \right) ~~~~~t\geq \gamma y
}
The condition on $t$ ensures that the integrand has a maximizer, and the asymptotic approximation is valid.  Intuitively, the ``wave" of influence has not reached $y$ before time $\gamma y$, and thus the expected state is still zero, assuming that function $s_0(t)$ only takes non-zero values for $t \ge 0$. 

\begin{figure}[t]
  \centering
  \includegraphics[width=1\columnwidth]{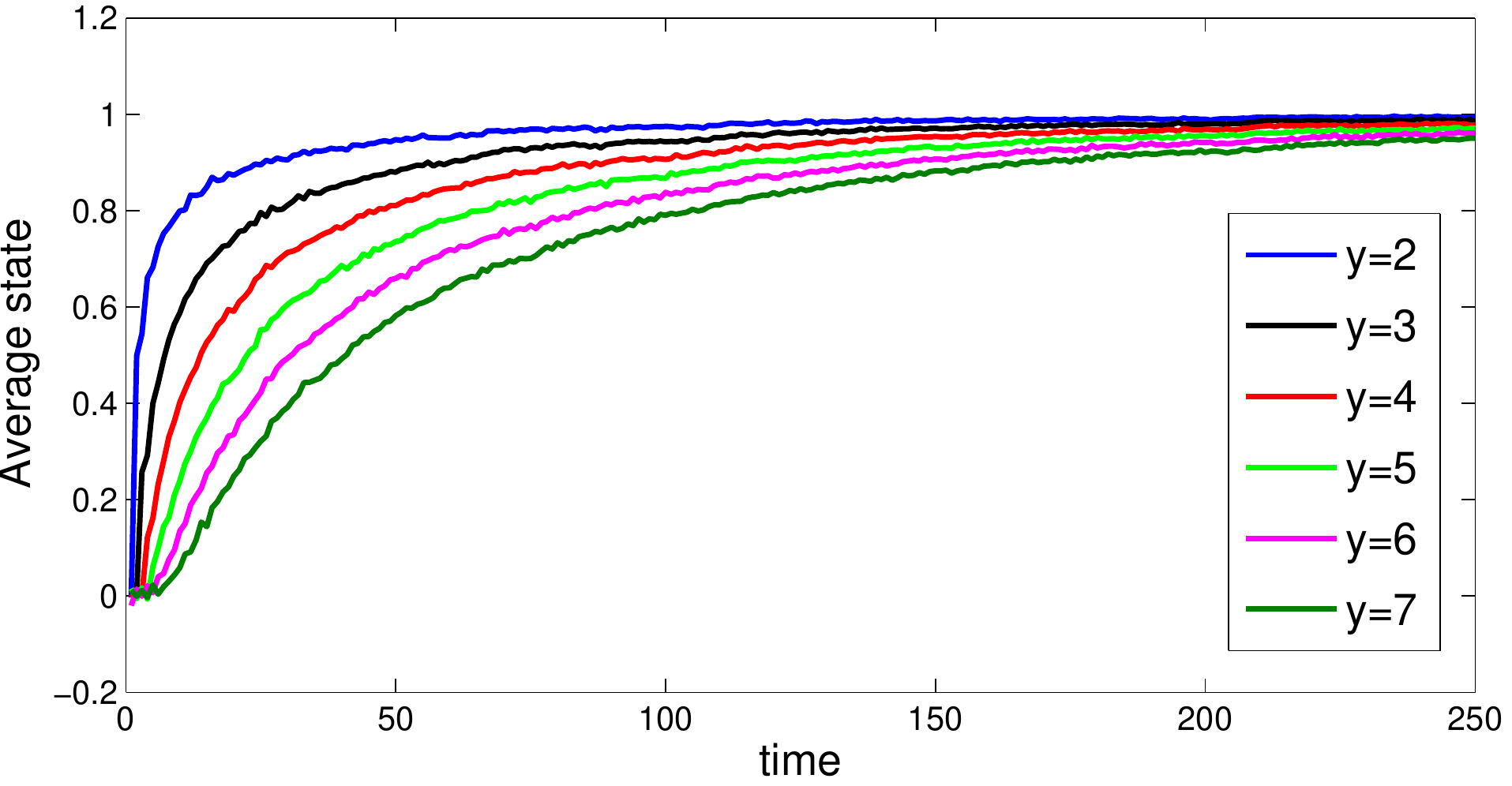}
  \caption[Figure ]%
  {Time evolution of the state on different sites on a semi-infinite structure when $\beta=1.2 \alpha q>\alpha q$ (thus the head node takes over), 
for $y=2$ (top one) to $y=7$. It is clear that, the closer to the head the class is located, the faster it complies. We simulated a chain of length 5000 to approximate the semi-infinite chain for the $y<250$ region, and averaged over 10000 Monte Carlo simulations. }
\label{inf_headwin}
\end{figure}

\begin{figure}[t]
  \centering
  \includegraphics[width=1\columnwidth]{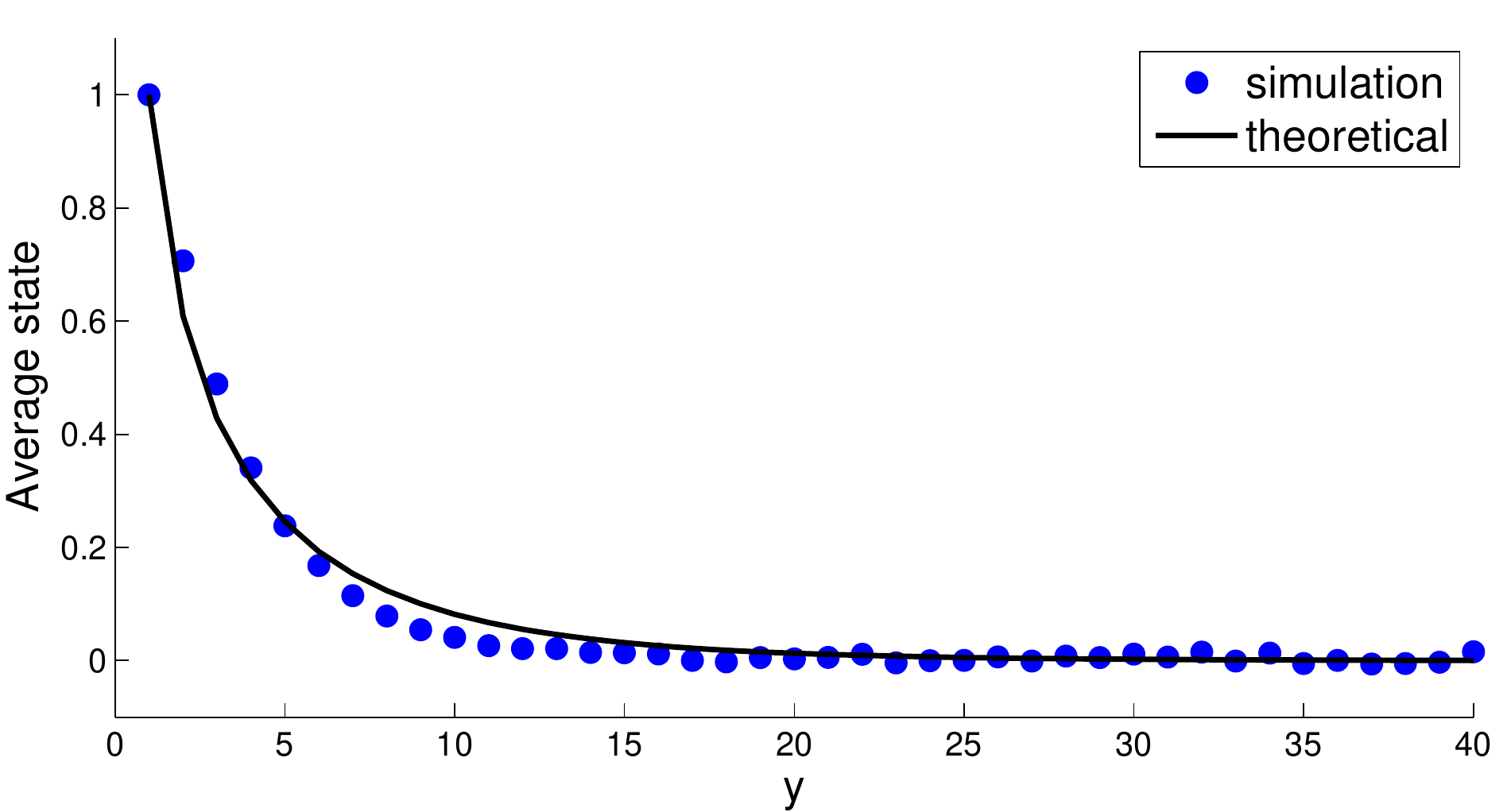}
  \caption[Figure ]%
  {Steady state values on a semi-infinite structure when $\beta=0.7 \alpha q<\alpha q$, thus the influence of the head node dies out. We simulated a chain of length 5000 to approximate the semi-infinite chain for the $y<250$ region. Monte Carlo simulations were added until the absolute value of the relative change in the mean state, averaged over all cites,  became less than $10^{-5}$.  }
\label{inf_headlose}
\end{figure}

For the problem at hand, the head node is fixed at state $s_{0}(t)=+1$, for all $t\geq 0$. This implies that the function $s_0(t)$ is the Heaviside step function $u(t)$, resulting in
\al{
s_y(t) \sim &\sqrt{\frac{|q\alpha-\beta|}{q \alpha + \beta + p}}~
\frac{\exp \left[ -\left( \frac{q\alpha-\beta}{2q\alpha}\right)y\right]}{\sqrt{y}} \nonumber \\
 &\times u\left(t-y \frac{q \alpha + \beta + p}{|q\alpha-\beta|}\right) 
.}
Depending on the sign of $(q \alpha - \beta) / \alpha$, the exponential factor will either grow or decay in $y$, giving rise to two different behaviors (see Figures~\ref{inf_headwin} and~\ref{inf_headlose}): 
\begin{enumerate}
\item When $q\alpha < \beta$ then $s_y(t) \rightarrow +1$ for all nodes (as illustrated in Figure~\ref{inf_headwin}).
\item When $q\alpha > \beta$ then the expected value of the spin decays in $y$, the grip of the head on the system debilitates the further one is from the head node (as illustrated in Figure~\ref{inf_headlose}).
\end{enumerate}

\subsection{Continuous Approximation: Finite Structure}
Next we study equation (\ref{sdot1}) for the case of a finite hierarchy, $y_{max}=L < \infty$. Let us define the constants
\eq{
\begin{cases}
 \displaystyle \frac{q\alpha}{q\alpha+\beta+p}  \equiv a \\ \\
\displaystyle   \frac{q\alpha -\beta}{q\alpha+\beta+p}  \equiv b.
\end{cases}
}
Using these, we rewrite (\ref{sdot1}) as
\eqq{ 
\rond{ s_y(t)}{t} = 
a  \frac{\partial^2  s_y(t) }{\partial y^2}+ b \rond{s_y(t)  }{y}
.}{sdot2}

The state of the head node is fixed at $s_0(t)=+1$. We turn our attention to the case  where there is complete disagreement between the head and the 
bottom-most class; i.e., the bottom-most class is fixed at $s_L(t)=-1$.  

\begin{figure}[t]
  \centering
  \includegraphics[width=1\columnwidth]{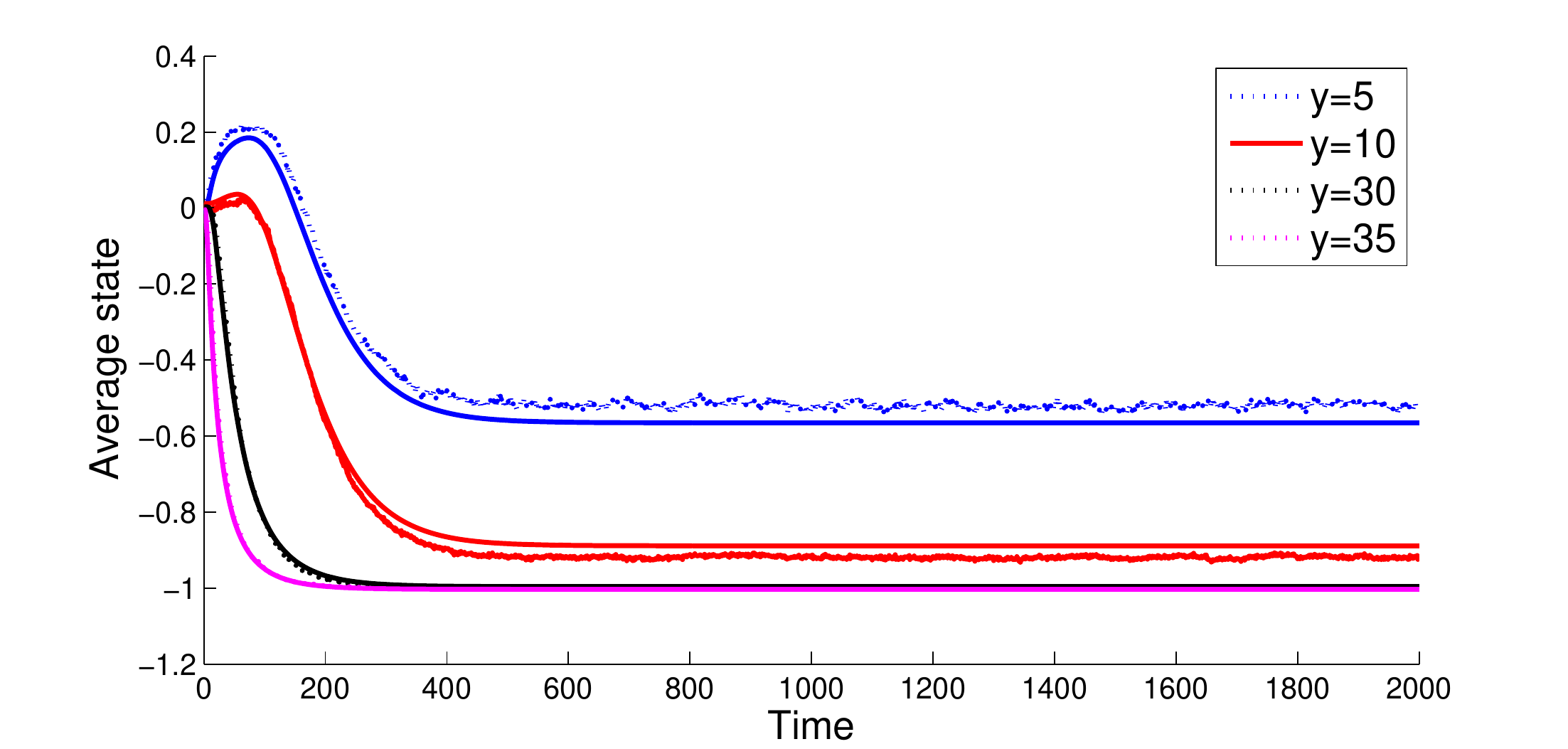}
  \caption[Figure ]%
  {Time evolution of states on a finite chain when $\beta=0.7 \alpha q<\alpha q$, implying the dominance of the universal class.
 The solid lines represent the theoretical prediction. The height of the structure is $L=40$ 
in this simulation. The two at the bottom ($y=30, 35$) are frozen at a state very close to $-1$, because of the proximity to the universal class at $y=L=40$. The results are averaged over $12000$ Monte Carlo simulations.}
\label{fig_fourier_1}
\end{figure}

\begin{figure}[t]
  \centering
  \includegraphics[width=1\columnwidth]{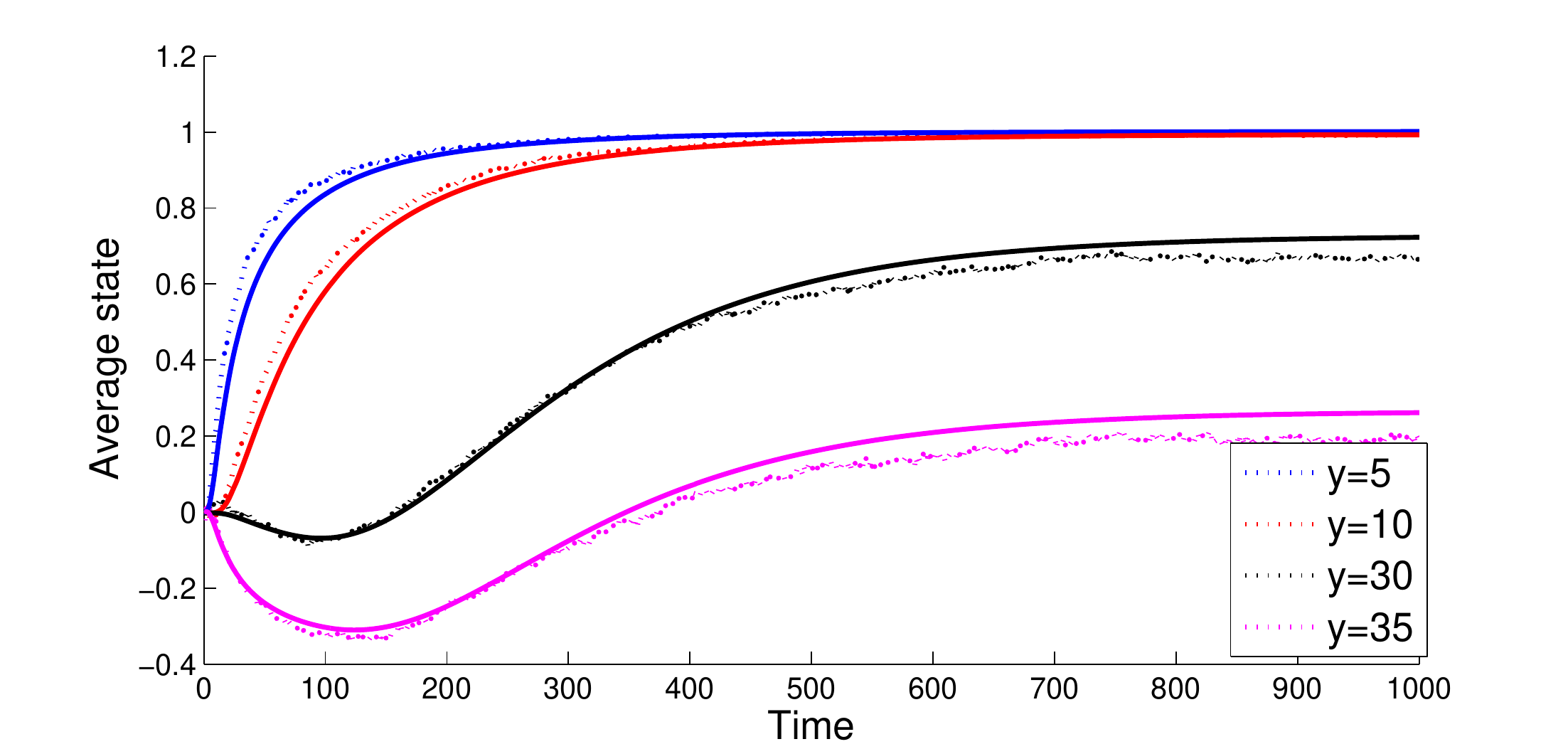}
  \caption[Figure ]%
  {Time evolution of states on a finite chain when $\beta=1.2 \alpha q>\alpha q$, whose ramification is the dominance of the head.
 The solid lines represent the theoretical prediction. The height of the structure is $L=40$. The one at the bottom corresponds to $y=35$, at the vicinity 
of the universal class, but frozen near zero rather than $-1$. The state of the top two ($y=5, 10$), neighboring the head, is almost exactly $+1$.  
in this simulation. The results are averaged over $12000$ Monte Carlo simulations. }
\label{fig_fourier_2}
\end{figure}

\begin{figure}[t]
  \centering
  \includegraphics[width=1\columnwidth]{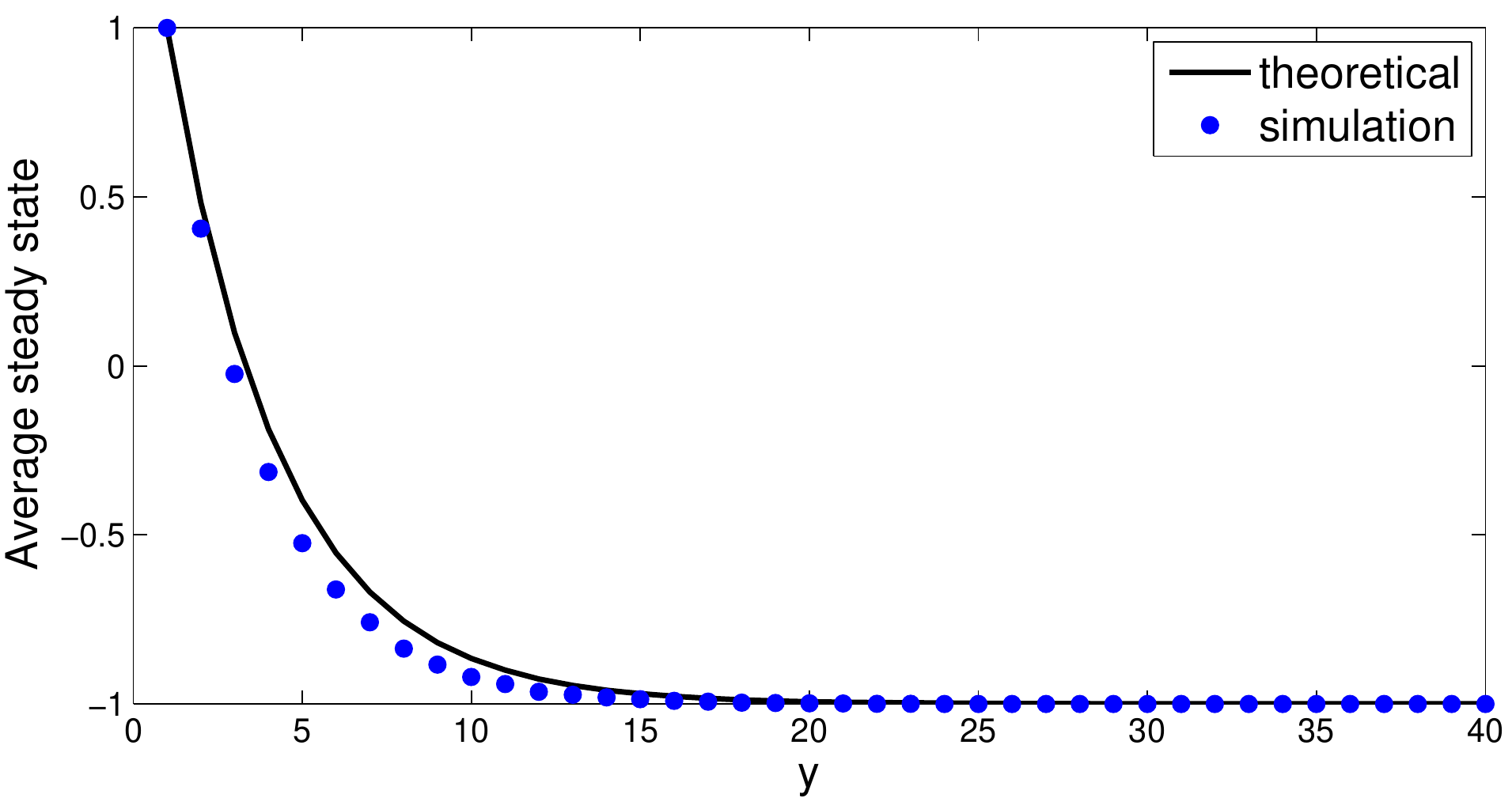}
  \caption[Figure ]%
  {Steady state on a finite chain with length $L=40$ when $\beta=0.7 \alpha q<\alpha q$ . Note that the majority is converted to the lead of the bottom class.  The results are averaged over $15000$ Monte Carlo simulations,  when  the  absolute value of the relative change in the mean state,  averaged over all cites, became less than $10^{-5}$. }
\label{fig_fourier_steady_1}
\end{figure}

\begin{figure}[t]
  \centering
  \includegraphics[width=1\columnwidth]{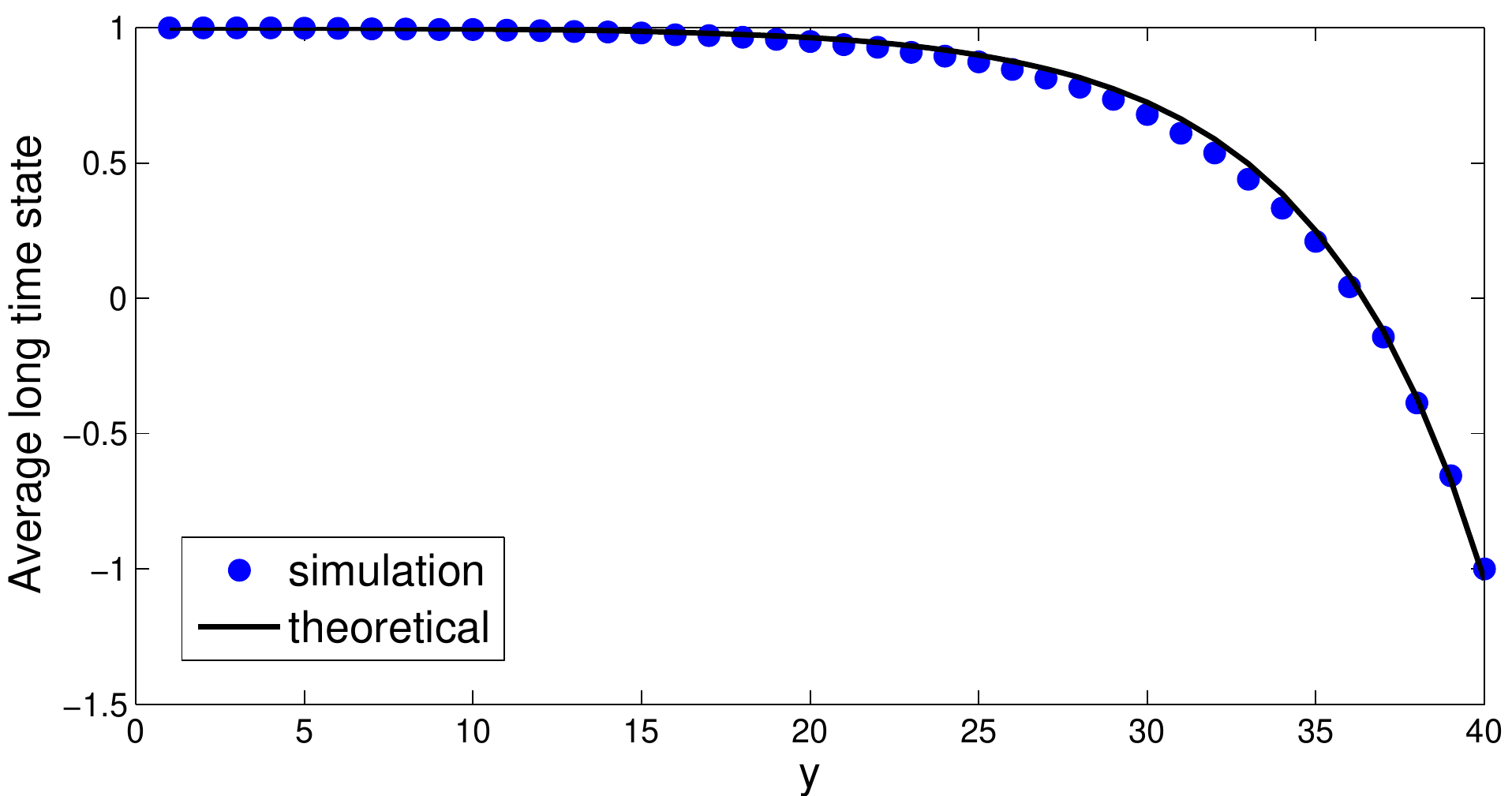}
  \caption[Figure ]%
  {Steady state on a finite chain with length $L=40$ when $\beta=1.2 \alpha q>\alpha q$. The head node prevails clearly. The results are averaged over $8000$ Monte Carlo simulations,  when  the  absolute value of the relative change in the mean state, averaged over all cites,  became less than $10^{-5}$. }
\label{fig_fourier_steady_2}
\end{figure}

With the substitution  $s_y(t)=\phi(y,t) \exp \left( \frac{-b}{2a} y -\frac{b^2}{4a} t \right) $, \eqref{sdot2} is recognizable as the standard homogeneous heat equation for $\phi(y,t)$. Thus we arrive at:
\all{
s_y(t) = &\frac{2\pi}{L^2} \sum_{n=1}^{\infty} 
n \sin \left( \frac{n \pi y}{L} \right) \frac{\left(1-\exp  \left[-\left( \frac{ n^2 \pi^2 a}{L^2}+\frac{b^2}{4a^2}\right) t\right]\right)}{ \frac{  n^2 \pi^2}{L^2}+\frac{b^2}{4a^2}}  \nonumber \\ 
&\times \left[ \exp \left( -\frac{by}{2a}\right) +(-1)^n \exp \left( \frac{b(L-y)}{2a}\right) \right]
,}{fourier1}
as depicted in Figures \ref{fig_fourier_1} and \ref{fig_fourier_2}. For the steady state, when the time derivative  vanishes, we have
\all{
\lim_{t \rightarrow \infty } s_y(t) &=  \frac{2\pi}{L^2}  \displaystyle \sum_{n=1}^{\infty} 
\Bigg\{n \displaystyle \sin \left( \frac{n \pi y}{L}  \right)
 \nonumber \\
& 
\times \frac{ \exp \displaystyle\left( -\frac{by}{2a}\right)+(-1)^n \displaystyle\exp \left( \frac{b(L-y)}{2a}\right)}{\displaystyle\frac{ \pi^2 n^2}{L^2}+\frac{b^2}{4a^2}} \Bigg\}
.}{sy_inf}
To simplify this further,  observe that the steady state can be expressed in a simpler form by noting that (\ref{sdot2}) reduces to 
\eq{ 
a  \frac{d^2}{dy^2}s_y + b \frac{d}{dy} s_y=0.}
When $s_L(t)=-1$ and $s_0(t)=1$, the solution is given by
\eqq{
\lim_{t\rightarrow \infty}s_y(t)=\frac{2 \exp\left(\frac{-b}{a}y\right)-1-\exp\left(\frac{-b}{a}L\right)}{1-\exp\left(\frac{-b}{a}L\right)}
.}{fourier_steady}
The expression in~\eqref{sy_inf} is the Fourier expansion of~\eqref{fourier_steady}, as shown in Appendix~\ref{app:equiv_1}. The results are illustrated in Figures~\ref{fig_fourier_steady_1} and~\ref{fig_fourier_steady_2}.

Note that, in the case of agreement between the two ends, we have
\eq{
\lim_{t\rightarrow \infty}s_y(t) =+1~~\forall y
,}
which signifies the importance of the role of the bottom class. 

\section{Second Model: Constant Fraction of Intra-class Links}
\label{sec:constantFraction}

In this section we suppose that the number of neighbors each node has at the same level is a constant fraction of the size of that level; this is in contrast to the model studied in the previous section where each node had exactly $p$ neighbors at the same level. Recall that there are $q^y$ nodes at level $y$. In this section, we model the weight assigned by nodes at level $y$ to the average opinion of nodes at that same level as $p_y = R q^y$ for some positive constant $R \le 1$. Structurally, this corresponds to having the connections from a node at level $y$ to each other node at level $y$ be present with probability $R$. Thus, the global structure is obtained by forming the hierarchy to obtain levels, and then creating an Erd\H{o}s-R\'{e}nyi (i.e., Bernoulli) random graph among the nodes within each level, for which the average intra-level degree is taken to be linear in the number of nodes at that level. 

The evolution of the expected values of the states becomes
\begin{eqnarray}
\dot{s}_y=&& -\left(\frac{q \alpha + \beta}{q \alpha + \beta + R q^y} \right) s_y  
+\left( \frac{\beta}{q \alpha + \beta +R q^y}\right) s_{y-1} \nonumber \\
&&+ \left( \frac{q \alpha}{q \alpha + \beta + R q^y}\right) s_{y+1}
.
\label{sdot3}
\end{eqnarray}

Let us first consider  the case of strong hierarchy, i.e.,  $q \alpha \ll \beta, p_y$. For $\alpha=0$, there is complete obedience to the superior class, and no weight is assigned to the inferior classes. For this case we have
\eq{
\partial_t s_y(t)=\left( \frac{-\beta}{\beta+ R q^y} \right) \partial_y s_y(t)
.}
If the head node is fixed at $s=+1$, which means $s_{0}(t)=u(t)$, then we have
\eq{
s_y(t)=u\left( t-y-\frac{R q^y}{\beta \ln q}\right)
,}
which implies that for long times eventually all levels conform to the head node. 

The other polar case would be where $q \alpha \gg \beta, p_y$ which gives
\eq{
\partial_t s_y(t)= \partial^2_{yy} s_y(t)+\partial_y s_y(t)
,}
whose solution for long times follows
\eq{ 
s_y(t) \sim \frac{e^{-y}}{\sqrt{y}} u(t-y)
,}
which decays in $y$.

For the steady state, (\ref{sdot2}) can be written as: 
\eq{
\left( \frac{q \alpha}{q \alpha + \beta + R q^y} \right) \frac{d^2}{dy^2}  s_y + \left( \frac{q \alpha - \beta}{q \alpha + \beta + R q^y} \right) \frac{d}{dy} s_y=0
,}
whose solution is of the form
\eq{
s_y=K_1 \exp \left( \frac{\beta-q\alpha}{q\alpha} y \right) + K_2
,}
where $K_1$ and $K_2$ are constants that are uniquely determined for given boundary conditions.

The solution in the case of $s_L(t)=-1$, $s_0(t)=1$ coincides with (\ref{fourier_steady}), which does not depend on the value of $R$. Also, if $s_L(t)=s_0(t)=+1$ for all $t$, then we have $\lim_{t \rightarrow \infty} s_y(t)=1$ for all $y$.  Note that:
\begin{enumerate}
 \item The value of $R$ does not affect the steady state, but only how fast the system reaches it;
\item The only determinant factor of conformity is the relative value of $\beta$ and $q \alpha$; if the latter is greater, the  state imposed by the head node decays,  and it takes over otherwise. \\  
\end{enumerate}

\section{General Steady-State Discrete Solution: Finite Hierarchy}
\label{sec:discrete}

In this section we consider the general case for intraclass connectivity. We also take $y$ to be discrete. Let the average degree of the class at level $y$ be $p(y)$. Then the evolution of the expected value of the state is
\begin{eqnarray}
\dot{s}_y(t)= -&&\left(\frac{q \alpha + \beta}{q \alpha + \beta + p(y)} \right) s_y(t) \nonumber \\
+ && \left( \frac{\beta}{q \alpha + \beta + p(y)}\right) s_{y-1}(t) \nonumber \\
+ && \left( \frac{q \alpha}{q \alpha + \beta + p(y)}\right) s_{y+1}(t)
.
\label{sdot4}
\end{eqnarray}

Now we focus on the steady state, thus we drop the time arguments. The steady state satisfies
\begin{eqnarray}
-&&\left(\frac{q \alpha + \beta}{q \alpha + \beta + p(y)} \right) s_y \nonumber \\
+&& \left( \frac{\beta}{q \alpha + \beta + p(y)}\right) s_{y-1} \nonumber \\
+&& \left( \frac{q \alpha}{q \alpha + \beta + p(y)}\right) s_{y+1}=0
,
\end{eqnarray}
or equivalently, 
\eqq{
-(\alpha q+\beta) s_y + \beta s_{y-1} + \alpha q s_{y+1} = 0
,}{sss}
which is independent of $p(y)$. We emphasize that, \emph{the form of $p(y)$ does not affect the steady state but merely how fast the system arrives there}.

Equation (\ref{sss}) is a linear difference equation with constant coefficients~\cite{bender}. Therefore its solution has the form $s_y=K_1 \lambda_1^y + K_2 \lambda_2^y$. The coefficients $K_1$ and $K_2$ are determined using the boundary conditions. The factors $\lambda_1$ and $\lambda_2$ are readily determined by plugging the ansatz $s_y=\lambda^y$ into~\eqref{sss}, which gives:
\eq{
\alpha q \lambda^2 - (\alpha q +\beta) \lambda + \beta=0
.}
The two solutions to this quadratic equation yield $\lambda_1, \lambda_2$:
\eq{
\begin{cases}
\lambda_1= 1 \\ 
\lambda_2= \displaystyle \frac{\beta}{\alpha q}.
\end{cases}
}
To solve equation~\eqref{sss}, we require two boundary conditions to extract $K_1$ and $K_2$. Thus we consider  three distinct cases (the calculation of~$K_1$ and~$K_2$ is given in Appendix~\ref{app:find_k}).

\subsection{First Case: $s_0(t)=+1$ and $s_L(t)=-1$}

\begin{figure}[t]
  \centering
  \includegraphics[width=1\columnwidth]{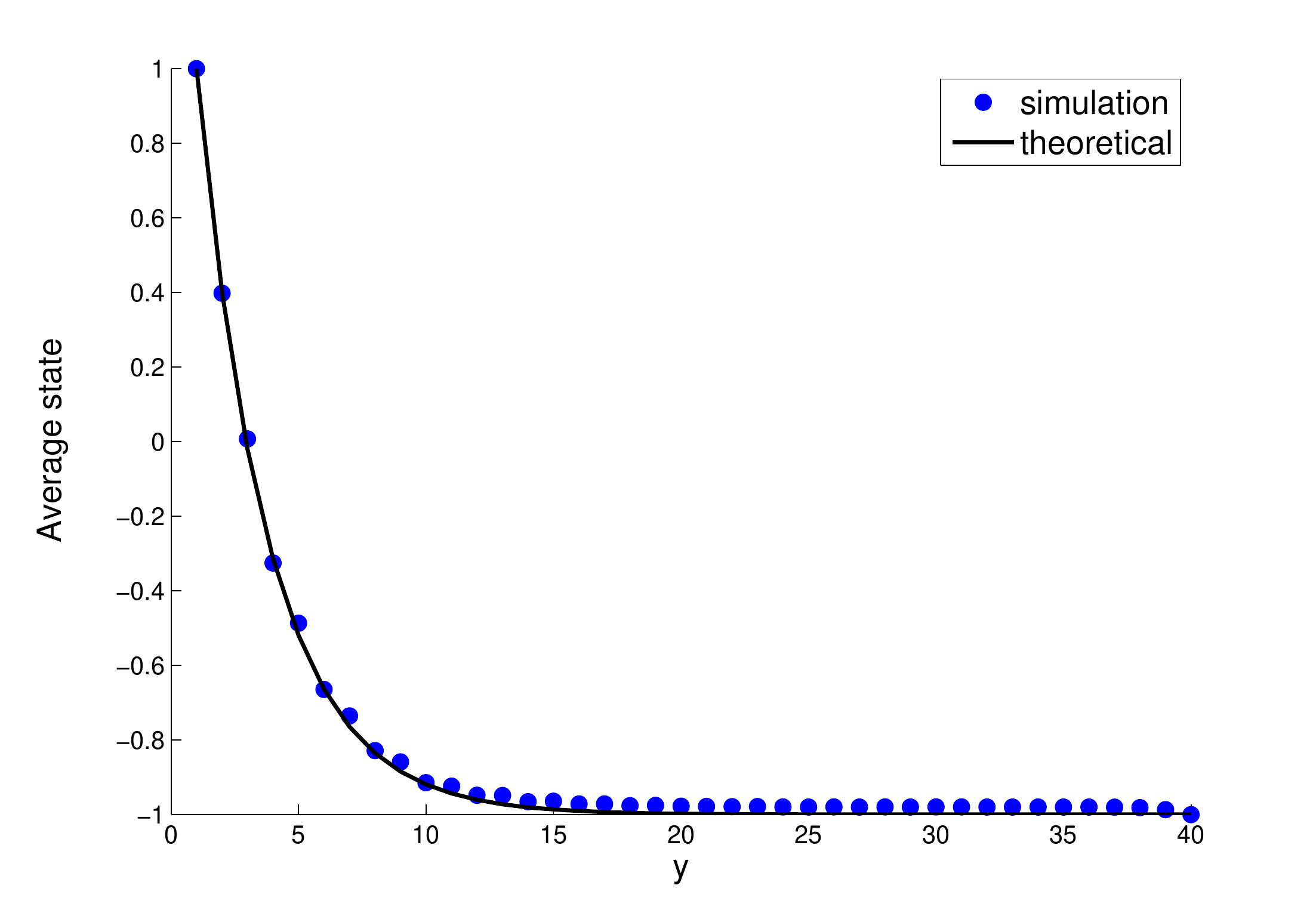}
  \caption[Figure ]%
  {Steady state on a finite chain with the bottom class opposing the head node, for the case $\beta=0.7 \alpha q< \alpha q$ for exponential $p(y)$. We averaged over 10000 Monte Carlo simulations, when  the  absolute value of the relative change in the mean state,  averaged over all cites, became less than $10^{-5}$. }
\label{p_change_endoppose_headlose}
\end{figure}

\begin{figure}[t]
  \centering
  \includegraphics[width=1\columnwidth]{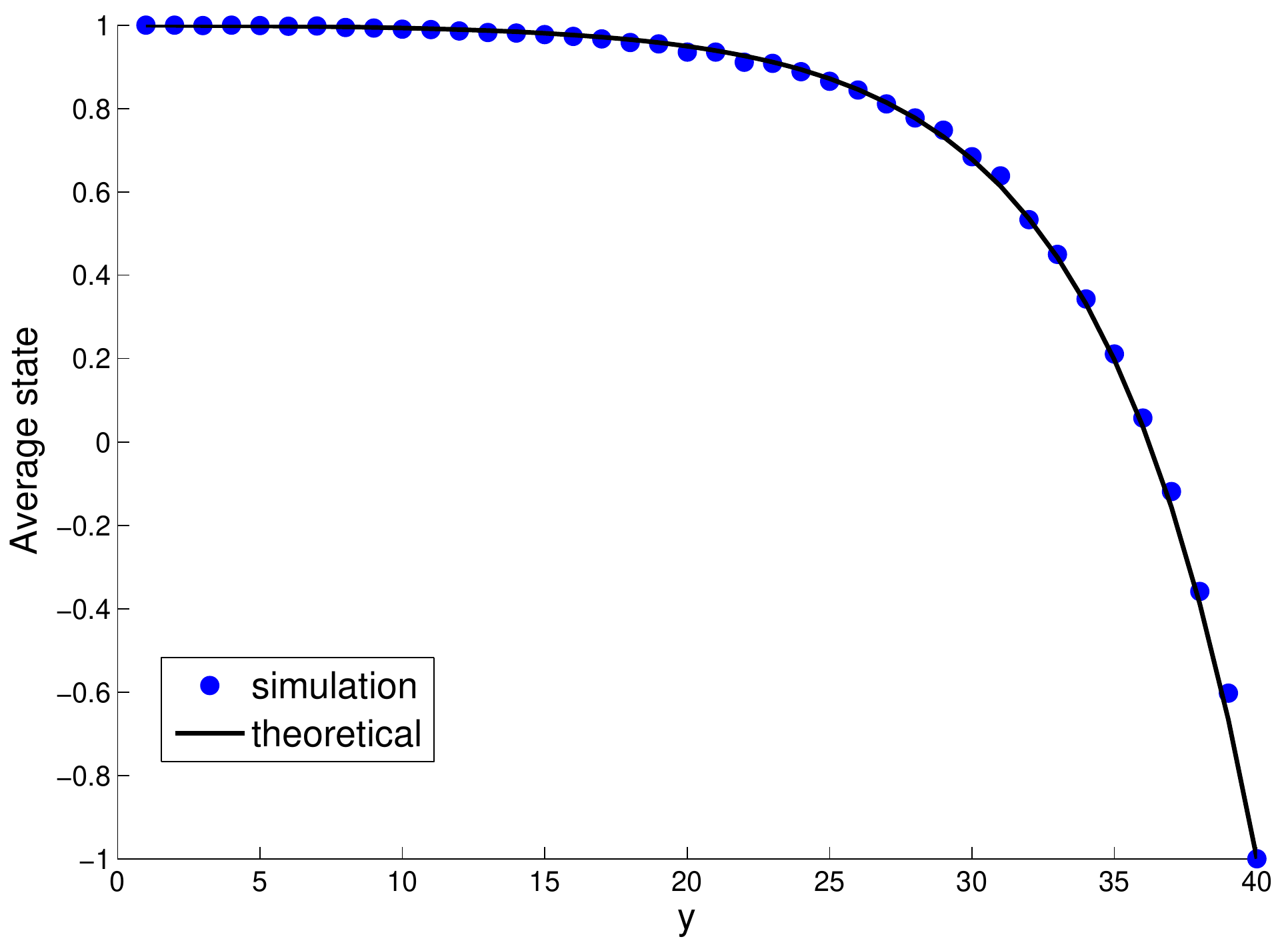}
  \caption[Figure ]%
    {Steady state on a finite chain with the bottom class opposing the head node, for the case $\beta=1.2 \alpha q>\alpha q$ for exponential $p(y)$. We averaged over 4000 Monte Carlo simulations, when  the  absolute value of the relative change in the mean state,  averaged over all cites, became less than $10^{-5}$. }
\label{p_change_endoppose_headwin}
\end{figure}

Suppose $s_0(t)=+1$ and $s_L(t)=-1$. Then the steady state value at level $y$ is 
\eq{
s_y=\frac{1+\left( \frac{\beta}{\alpha q}\right)^L-2\left( \frac{\beta}{\alpha q}\right)^y}{\left(\frac{\beta}{\alpha q}\right)^L-1} 
,}
which is a transition between the two states. The majority, across the entire network, is held by the head or the bottom depending on whether $\beta$ is bigger or smaller than $\alpha q$ respectively. Figures
\ref{p_change_endoppose_headlose} and \ref{p_change_endoppose_headwin} are depictions of these two possibilities for the case where $p(y)$ varies exponentially 
with $y$, meaning that a constant fraction of neighbors from each class are linked. 

Now let us calculate the average state among all nodes. Denote the number of nodes at level~$y$ by~$n_y$, and let $m$ denote the mean opinion across the entire network. We want to find
\eq{
m \stackrel{\text{def}}{=} \sum_{y=0}^L n_y s_y = \sum_{y=0}^L q^y s_y.
}
As shown in Appendix~\ref{app:find_k}, with $r = \beta / (\alpha q)$, in this case we have
\eq{
m=\displaystyle \frac{(r^L+1)}{(r^L-1) } - 
\displaystyle  \frac{2 \big[  (rq)^{L+1} -1 \big] (q-1) }{(q^{L+1}-1)(rq-1) (r^L-1)}
.}
The sign of the average state determines the collective inclination of the system, i.e., towards which state is the system leaning on average. In Appendix~\ref{app:find_k} we show that, as long as ${\alpha,\beta>0}$, the following is true: 

\emph{When the bottom-most class is fixed at the state~${s=-1}$, then:
\begin{enumerate}
\item If~$q>1$, then the mean state is negative, regardless of the values of~${\beta,\alpha}$. 
\item If~$q=1$, then the sign of the mean state is the same as the sign of~${\beta-\alpha}$.
\end{enumerate}
}

\subsection{Second Case: $s_0(t)=+1$ are Class $L$ is Influenceable}

\begin{figure}[t]
  \centering
  \includegraphics[width=1\columnwidth]{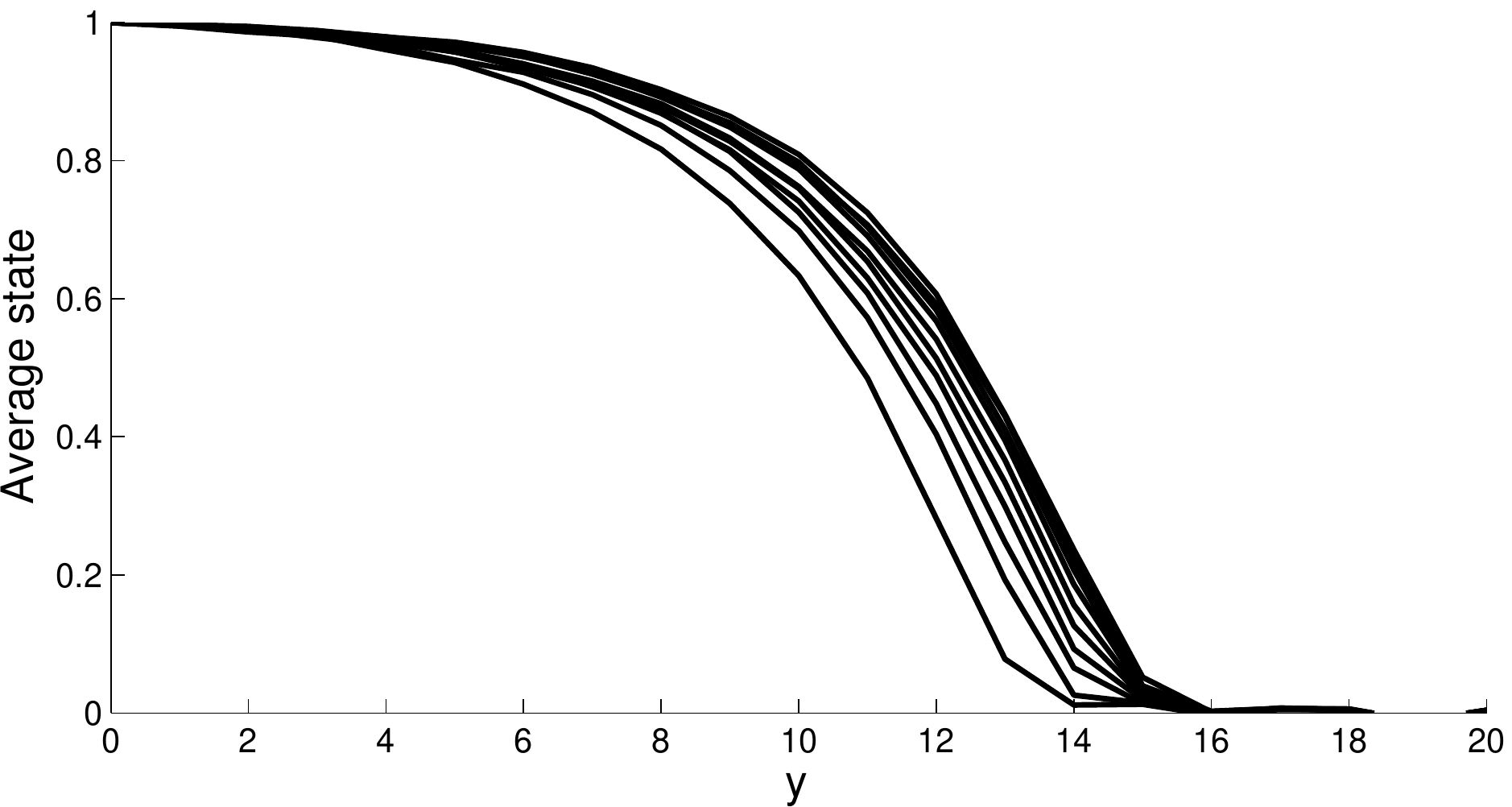}
  \caption[Figure ]%
  {State of the sites on an open-ended finite chain for different times, for 
the case $\beta=1.2\alpha q> \alpha q$. the left-most being $t=10$ and the rightmost $t=20$. The "wave" of influence is moving to the right. The results are averaged over 10000 Monte Carlo simulations. }
\label{p_change_endfree_headwin}
\end{figure}

Next, suppose that $s_0(t)=+1$ and the bottom class are influenceable, i.e., they follow the dynamics in (\ref{sdot4}), implying imitation of the upper class. In this case we have
\eq{
s_y=  \displaystyle  \frac{r^{L+1}-r^y}{r^{L+1}-1}
.}
Figures \ref{p_change_endfree_headwin} and \ref{p_change_endfree_headlose} illustrate this case. The average state is calculated in Appendix~\ref{app:open_end}. The result is: 
\eqq{
 \displaystyle m= \displaystyle \frac{r^{L+1}}{r^{L+1}-1} -  \displaystyle  \frac{\big[ (qr)^{L+1}-1\big] (q-1) }{(qr-1)(q^{L+1}-1)(r^{L+1}-1)}
.}{m_case2}
Taking the average over all nodes, we find that: 

\emph{For when the bottom-most class is influenceable, the mean state of the system is positive, regardless of the value of~$r$ and~$q$.}

\subsection{Third Case: $s_0(t) = +1$ and $s_L(t) = +1$}

Finally, suppose that both $s_0(t)=+1$ and $s_L(t)=+1$. In this case, $\lim_{t\rightarrow \infty} s(y)=1$ for all $y$, irrespective of values of $\beta, \alpha$ and $q$. 

The implication of the last case is straightforward. The head node does not need to exert influence on the middle classes to achieve its ends; rather, getting the bottom-most class aligned suffices to dominate the entire hierarchy. This signifies the importance of media, or any other means by which this alignment could be obtained.

\begin{figure}[t]
  \centering
  \includegraphics[width=1\columnwidth]{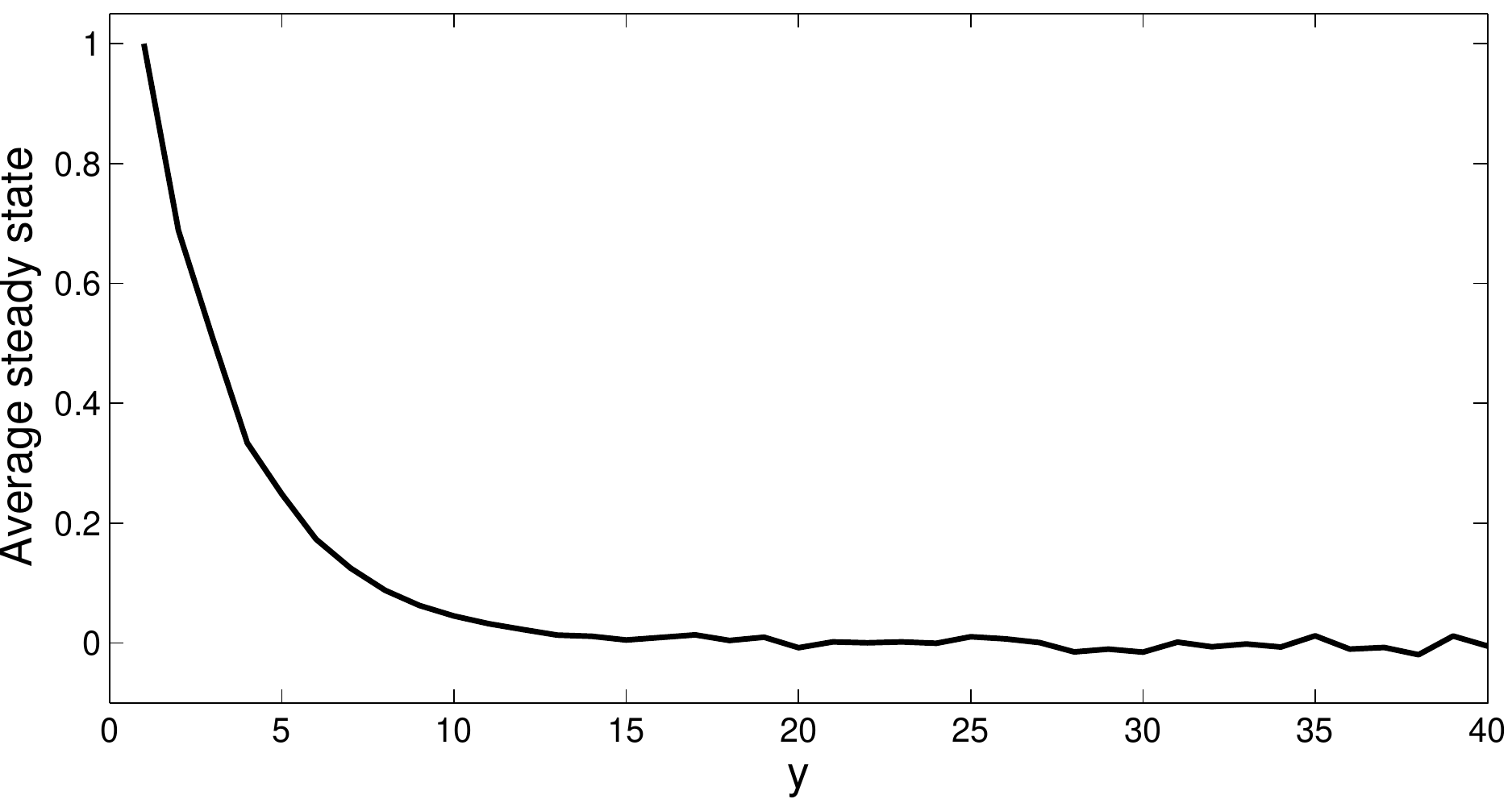}
  \caption[Figure ]%
  {Steady state on a  finite chain when $\beta=0.7\alpha q<\alpha q$. The bottom node is free, which means that it follows 
the class above it. We averaged over 21000 Monte Carlo simulations, when  the  absolute value of the relative change in the mean state,  averaged over all cites, became less than $10^{-5}$.}
\label{p_change_endfree_headlose}
\end{figure}

\section{Summary and Discussion} \label{sec:summary}

We studied the dynamics of binary states, for nodes  on a pyramid-shaped hierarchical graph. The head node was given a fixed state, trying to impose it on the whole network. 
The conditions that facilitate or hinder the dominance of the head node were found.
 If classes assign more weight to classes below them, dominance of the head node is hampered. It was observed that in the case 
of discord between the head node and the bottom class in a finite hierarchy, one of them wins the majority. Another result 
  was that, the density of the intra-class links does not affect the steady state, but merely determines how quickly the system reaches it. 
Also, if the head node and the bottom-most class are fixed at the same state, the whole hierarchy will eventually comply, under any conditions 
for the inter/intra-class links. 

Possible extensions of this work include using a more realistic topology for the underlying graph to model the social structure, considering continuous and/or multidimensional states, deploying more than one  classification attribute (e.g., Weberian class theory \cite{coser}), and  having level-dependent ratios for weights assigned to lower and higher classes. Also, one can examine the effects of adding sparse long-range interactions to the hierarchy. Another modification would be adding fractions of stubborn agents within the middle classes and studying their effect on the final state. Also, one can add exogenous influence to the whole system, in the form of an external field, with different exposure degrees for different classes. This can model the effect of mass media.


\appendix

\section{Proof of the Equivalence of~\eqref{sy_inf} and~\eqref{fourier_steady}} \label{app:equiv_1}
Let us begin with~\eqref{fourier_steady}, which is repeated here for easy reference: 
\eq{
\lim_{t\rightarrow \infty}s_y(t) = \frac{2 \exp\left(\frac{-b}{a}y\right)-1-\exp\left(\frac{-b}{a}L\right)}{1-\exp\left(\frac{-b}{a}L\right)}
.}
Let us factor out~${\exp\left( {\frac{-by}{2a}}\right) }$, to obtain
\all{
\lim_{t\rightarrow \infty}&s_y(t)  = \displaystyle  \frac{ \exp\left(\frac{-b}{2a}y\right)}{1-\exp\left(\frac{-b}{a}L\right)} \nonumber \\
&  \times \Bigg\{ 2 \exp\left(\frac{-b}{2a}y\right)-\bigg[ 1+\exp\left(\frac{-b}{a}L\right)\bigg]  \exp\left(\frac{+b}{2a}y\right) 
\Bigg\}
.}{chi_s}
Now let us define 
\al{
\chi(y) & \stackrel{\text{def}}{=} 
 2 \exp\left(\frac{-b}{2a}y\right)-\bigg[ 1+\exp\left(\frac{-b}{a}L\right)\bigg]  \exp\left(\frac{+b}{2a}y\right) 
.}
We want to find the Fourier representation of this function, in the following form
\eq{
\chi(y)=\sum_n a_n \sin \left(\frac{n \pi}{L}y \right) .
}
The coefficients~$a_n$ are calculated as follows:
\eq{
a_n= \frac{2}{L} \int_{y=0}^L \chi(y) \sin \left(\frac{n \pi}{L}y \right)
.}
We need the following integrals, which are evaluated  by twice integration by parts: 
\al{
\begin{cases}
&\displaystyle \int^y \displaystyle \exp\left(\frac{-b}{2a}y'\right)  \sin \displaystyle \left(\frac{n \pi}{L}y' \right)  dy'= \nonumber \\
 &\displaystyle \exp\left(\frac{-b}{2a}y\right)  \frac{\frac{-b}{2a}  \sin \displaystyle \left(\frac{n \pi}{L}y\right)  - \frac{n\pi}{L} \cos \displaystyle \left(\frac{n \pi}{L}y \right) }{\left( \frac{b^2}{4a^2}\right)+ \left( \frac{n^2 \pi^2}{L^2}\right)}  
\\ \\ \\
&\displaystyle \int^y \displaystyle \exp\left(\frac{+b}{2a}y'\right)  \sin \displaystyle \left(\frac{n \pi}{L}y' \right)  dy'= \nonumber \\
 &\displaystyle \exp\left(\frac{+b}{2a}y\right)  \frac{\frac{b}{2a}  \sin \displaystyle \left(\frac{n \pi}{L}y \right)  - \frac{n\pi}{L} \cos \displaystyle \left(\frac{n \pi}{L}y \right) }{\left( \frac{b^2}{4a^2}\right)+ \left( \frac{n^2 \pi^2}{L^2}\right)} 
\end{cases}
.}
The integration is from 0 to $L$. The~$\sin$ terms vanish at both ends, and we get
\al{
\begin{cases}
&\displaystyle \int_{y'=0}^L \displaystyle \exp\left(\frac{-b}{2a}y'\right)  \sin \displaystyle \left(\frac{n \pi}{L}y' \right)  dy'= \nonumber \\
 &\displaystyle \left( \frac{n \pi}{L}\right) \displaystyle \frac{ 1-(-1)^n   \exp \left( \frac{-bL}{2a}\right)}{\left( \frac{b^2}{4a^2}\right)+ \left( \frac{n^2 \pi^2}{L^2}\right)}  
\\ \\ \\
&\displaystyle \int_{y'=0}^L \displaystyle \exp\left(\frac{+b}{2a}y'\right)  \sin \displaystyle \left(\frac{n \pi}{L}y' \right)  dy'= \nonumber \\
 &\displaystyle  \left( \frac{n \pi}{L}\right)  \displaystyle  \frac{ 1-(-1)^n \exp \left( \frac{+bL}{2a}\right)}{\left( \frac{b^2}{4a^2}\right)+ \left( \frac{n^2 \pi^2}{L^2}\right)} 
\end{cases}
.}
Using these, the Fourier coefficients become
\al{
a_n&= \frac{4 n \pi}{L^2} \displaystyle \frac{ 1-(-1)^n  \exp \left( \frac{-bL}{2a}\right)}{\left( \frac{b^2}{4a^2}\right)+ \left( \frac{n^2 \pi^2}{L^2}\right)}  
\nonumber \\
&-\frac{2n\pi}{L^2}\bigg[ 1+\exp\left(\frac{-b}{a}L\right)\bigg]\displaystyle \frac{ 1-(-1)^n  \exp \left( \frac{+bL}{2a}\right)}{ \displaystyle \left( \frac{b^2}{4a^2}\right)+  \displaystyle \left( \frac{n^2 \pi^2}{L^2}\right)} 
,}
which simplifies to
\al{
a_n =& \frac{2n\pi}{L^2}  \displaystyle   \frac{1-\exp\left(\frac{-b}{a}L\right)-(-1)^n \exp\left(\frac{-b}{2a}L\right) + (-1)^n \exp\left(\frac{b}{2a}L\right) }
{\left( \frac{b^2}{4a^2}\right)+ \left( \frac{n^2 \pi^2}{L^2}\right)} \nonumber \\
&=  \displaystyle  \frac{2n\pi}{L^2}  \displaystyle   \frac{\bigg[  1-\exp \left(\frac{-b}{a}L\right)\bigg]   \bigg[  1+(-1)^n \exp  \left(\frac{b}{2a}L\right)   \bigg] }
{ \displaystyle \left( \frac{b^2}{4a^2}\right)+  \displaystyle \left( \frac{n^2 \pi^2}{L^2}\right)}
.}
Substituting this expression to represent~$\chi(y)$ in~\eqref{chi_s} is Fourier series form, we arrive at
\al{
&\lim_{t\rightarrow \infty} s_y(t)  =  
\sum_n \sin \left( \frac{n \pi}{L} \right)   \nonumber \\
&\times\frac{2n\pi}{L^2}
\frac{ \bigg[   \exp\left(\frac{-b}{2a}y\right) +(-1)^n \exp  \left(\frac{b}{2a}(L-y)\right)   \bigg] }
{ \displaystyle \left( \frac{b^2}{4a^2}\right)+  \displaystyle \left( \frac{n^2 \pi^2}{L^2}\right)}
,}
which is identical to~\eqref{sy_inf}, as we wanted to illustrated.

\section{Complete Solution of~\eqref{sss} for  the Boundary Condition of Fixed Opposite Ends, and Discussion on the Average State} \label{app:find_k}

Using the ansatz~$s_y=\lambda^y$ in the homogeneous linear difference equation
\eq{
-(\alpha q+\beta) s_y + \beta s_{y-1} + \alpha q s_{y+1} = 0
,}
we get the characteristic equation: 
\eq{
\alpha q \lambda^2 - (\alpha q +\beta) \lambda + \beta=0
.}
The roots are:
\eq{
\begin{cases}
\lambda_1= 1 \\ 
\lambda_2= \displaystyle \frac{\beta}{\alpha q}.
\end{cases}
}
Let us define
\eq{
r \stackrel{\text{def}}{=} \displaystyle \frac{\beta}{\alpha q}
.}
The general solution is of the form~$s_y=K_1 \lambda_1^y + K_2 \lambda_2^y$. We have the boundary conditions
${s_0=1}$ and ${s_L=-1}$. The first condition gives: 
\eq{
K_1+K_2=1.}
The second one gives
\eq{K_1  + K_2 r^L= -1}
Solving for~$K_1$ and~$K_2$, we get
\eq{
\begin{cases}
K_1=  \displaystyle  \frac{1+r^L}{r^L-1}  \\
K_2= \displaystyle   \frac{-2}{r^L-1}
\end{cases}
.}
So the solution is
\eq{
s_y= \frac{1+r^L-2r^y}{r^L-1}
.}

To take the average state, we first have to find the following sum:
\all{
&\sum_{y=0}^L \frac{(1+r^L)q^y-2(rq)^y}{r^L-1} \nonumber \\
&=\frac{(r^L+1)}{r^L-1} \sum_{y=0}^L q^y - \frac{2}{r^L-1} \sum_{y=0}^L (rq)^y \nonumber \\
&= \displaystyle \frac{(1+r^L)(q^{L+1}-1)}{(r^L-1) (q-1)} - 
\displaystyle  \frac{2 \big[  (rq)^{L+1} -1 \big] }{(r^L-1)(rq-1)}.
}{temp_sum_1}
Note that the total number of nodes is
\eq{
\sum_{y=0}^L q^y =\frac{q^{L+1}-1}{q-1}
.}
Dividing~\eqref{temp_sum_1} by the number of nodes, yields the average state over all nodes
\eqq{
m=\displaystyle \frac{(r^L+1)}{(r^L-1) } - 
\displaystyle  \frac{2 \big[  (rq)^{L+1} -1 \big] (q-1) }{(q^{L+1}-1)(rq-1) (r^L-1)}
.}{m_temp_1}
Now let us show that for any~$q>1$, the average state is negative, regardless of~$L$ and~$r$. Negative average state is indicative of the dominance of the bottom-most class. 

First let us consider the case of~$r>1$, that is, the value of~$\beta$ being greater than~$\alpha q$. In this case,   since~$r^L-1$ is positive, we have to show that the following is true:
\eq{
\displaystyle (r^L+1) - 
\displaystyle  \frac{2 \big[  (rq)^{L+1} -1 \big] (q-1) }{(q^{L+1}-1)(rq-1)} <0
.}
It is equivalent to
\eqq{
2 \big[  (rq)^{L+1} -1 \big] (q-1) - (q^{L+1}-1)(rq-1) (r^L+1)  >0
,}{f_equiv}
which, after rearranging the terms and separating different powers of~$r$, transforms into
\al{
&r^{L+1} \left( q^{L+2} - 2q^{L+1}+q \right) 
+ r^L \left( q^{L+1}-1 \right)  \nonumber \\
&- r q \left( q^{L+1}-1 \right)
+ \left( q^{L+1}+1-2q \right) >0
.}
Let us define
\all{
f(q,r) \stackrel{\text{def}}{=} &r^{L+1} \left( q^{L+2} - 2q^{L+1}+q \right) 
+ r^L \left( q^{L+1}-1 \right)  \nonumber \\
&- r q \left( q^{L+1}-1 \right)
+ \left( q^{L+1}+1-2q \right) 
}{f1_temp1}
This can  also be rearranged as follows:
\all{
f(q,r) =  &q^{L+2} \left( r^{L+1} - r \right) 
+ q^{L+1} \left( r^L-2r^{L+1}+1 \right)  \nonumber \\
&+ q \left( r^{L+1}+r-2 \right)
+ \left(1-r^L \right) 
.}{f1_temp2}
The objective is to show that~$f(r,q)$ is positive for all integers~${q>1}$ and all~${r>1}$. 
Now we show that~$f(q,1)=0$,  that the first  derivative of~$f(\cdot)$ with respect to~$r$  at~${r=1}$ is nonnegative and that the  second derivative  of~$f(\cdot)$ is nonnegative for all~${r\geq 1}$ and all integers~${q>1}$. From~\eqref{f1_temp2} is is clear that:
\eqq{
f(q,1)=0.
}{f1_at_1}

Now we take the first derivative of~\eqref{f1_temp1}. We get:
\all{
\rond{f(q,r)}{r}=  & (L+1)r^{L} \left( q^{L+2} - 2q^{L+1}+q \right) \nonumber \\
&
+ Lr^{L-1} \left( q^{L+1}-1 \right)  -  q \left( q^{L+1}-1 \right)
.}{f1_deriv}
Evaluating this at~$r=1$, we get
\all{
\left. \rond{f(q,r)}{r}\right|_{r=1}=  & (L+1)\left( q^{L+2} - 2q^{L+1}+q \right) \nonumber \\
&
+ L\left( q^{L+1}-1 \right)  -  q \left( q^{L+1}-1 \right)
.}{f1_deriv2}
To confirm that this is positive for all~${q\geq2}$, we first show that its value at~${q=2}$ is positive, and then show that its derivative with respect to~$q$ is always nonnegative. We have: 
\al{
\left. \rond{f(q,r)}{r}\right|_{q=2,r=1}=  & (L+1)\left( 2^{L+2} - 2^{L+2}+2 \right) \nonumber \\
&
+ L\left( 2^{L+1}-1 \right)  -  2 \left(2^{L+1}-1 \right) \nonumber \\
&= L+2+2^{L+1}(L-2)
,}
which  is positive  for all valid values of~$L$, which are integers greater than zero (note that  at~${L=1}$ the value of the derivative is 1, and for larger values of~$L$, all terms are positive). Now that we have shown the value of the derivative at~${q=2}$ is positive, we take the derivative to verify that it is monotonically increasing for greater values of~$q$. Taking the derivative of~\eqref{f1_deriv2}, we have: 
\begin{gather}
~~~~\left. \displaystyle \rondx{f(q,r)}{r}{q}\right|_{r=1} = (L+1)\bigg[ (L+2)q^{L+1} - 2(L+1)q^{L}+1 \bigg] \nonumber \\
+ L  (L+1)q^{L}   -   \bigg[ (L+2)q^{L+1}-1 \bigg] \nonumber \\
~~~~= q^{L+1}L (L+2) + q^L (L+1)(L-2L-2)+(L+2) \nonumber \\
= (L+2) \bigg[ L q^{L+1}-(L+1)q^L+1    \bigg] 
.
\end{gather}
To show that this is nonnegative, is suffices to show that
\eq{
Lq^{L+1} > (L+1) q^L
,}
which is equivalent to
\eq{
q\geq L+\frac{1}{L}
,}
which is true, since we have~${q \geq 2}$ and $L$ is at least one. 

The last step is to show that the second derivative of~$f(\cdot)$ with respect to~$r$ is nonnegative. Taking the derivative of~\eqref{f1_deriv}, we get
\all{
\rondd{f(q,r)}{r}=  & L(L+1)r^{L-1} \left( q^{L+2} - 2q^{L+1}+q \right) \nonumber \\
&
+ L(L-1)r^{L-2} \left( q^{L+1}-1 \right)  
.}{f1_deriv}
The second term is nonnegative. Now we prove that the first term is also nonnegative for all~${q \geq 2}$. Note that the following is true:
\al{
 q^{L+2} - 2q^{L+1}+q = q^{L+1} (q-2) + q
.}
Each term is nonnegative for~${q\geq 2}$, hence the second derivative is always positive. Hence  for the case of~${r>1}$, we have proved that~${f(q\geq 2 ,r)}$ is always positive, so~$m$ is negative.

The second case we consider is when~${r<1}$ and~${rq>1}$. Looking back at~\eqref{m_temp_1}, we now have to prove that~${f(q,r)}$ is nonpositive in this range of~$r$. Note that we have already proven that~${f(\cdot)}$ is convex in~$r$, and it is zero at~${r=1}$. From~\eqref{f_equiv} it is readily seen that~${r=\frac{1}{q}}$ is also a root. A convex function can at most have two zeros, which we have found. In between, it is nonpositive, as we sought to prove. 

The final case is when~${r<1}$ and~${rq<1}$. In this case, from~\eqref{m_temp_1} we know that~${f(\cdot)}$ must be nonnegative in this range. With a reasoning similar to the previous case, we know that a convex function is nonnegative on the left hand side of its left root, as we desired~${f(\cdot)}$ to be. 

In sum, we have showed that~\eqref{m_temp_1} is nonpositive, regardless of~$r$, for all integers~${q>1}$. 

The case of~${q=1}$ is peculiar. Note that the geometric series summation which resulted in~\eqref{m_temp_1} must be modified, since this time the total population is~${L+1}$. The mean state is
\eqq{
m=\displaystyle \frac{(r^L+1)}{(r^L-1) } - 
\displaystyle  \frac{2 \big[  r^{L+1} -1 \big]} {(r^L-1) (r-1)(L+1)}
.}{m_temp_q1}
Taking the common denominator, we have
\eq{
m=\displaystyle \frac{(r^L+1)(r-1)(L+1) -  2 \big[  r^{L+1} -1 \big]} {(r^L-1) (r-1)(L+1)}
.}
Note that tis becomes zero for~${r=1}$. We intend to show that this is positive for~${r>1}$ and 
negative in the range~${0<r<1}$. The denominator is always positive (both factors flip sign at~${r=1}$). Let us denote the numerator by~$g(r)$. We have
\eq{
\begin{cases}
g(0)=1-L \leq 0 \\
g(1)= 0
.
\end{cases}
}
Taking the derivative, we have
\al{
\frac{d}{dr} g(r)&=(L+1) \bigg[ L r^{L-1} (r-1)+r^L+1    \bigg] - 2 (L+1) r^L  \nonumber \\
&= (L+1) \bigg[   L r^{L-1} (r-1)+1    -  r^L  \bigg] 
.}
Evaluating the derivative at points~${r=0}$ and~${r=1}$, we get
\eq{
\begin{cases}
\left. \displaystyle  \frac{d}{dr} g(r) \right|_{r=0}=L+1 > 0 \\ \\
\left. \displaystyle \frac{d}{dr} g(r) \right|_{r=1}= 0
.
\end{cases}
}
The second derivative is
\al{
&\frac{d^2}{dr^2} g(r)=(L+1) \Bigg[   L(L-1) r^{L-2} (r-1)    \nonumber \\
& +Lr^{L-1} -L r^{L-1}  \Bigg] = L (L^2-1) (r-1) r^{L-2}
,}
whose roots are only~${r=0,1}$. Also note that the second derivative is strictly negative in the range~${0<r<1}$, which is indicative of its concavity. We recognize the following behaviors: 
\begin{enumerate}
\item $r>1$: since the derivative is zero at~${r=1}$ and is positive for all~${r>1}$, the function~${g(\cdot)}$  will have a greater value in this range, than at~${r=1}$. Since the function, like its derivative, is zero at this point, so it will have a positive value for all~${r>1}$. 

\item $0<r<1$: the function has a negative value at~${r=0}$ and its vicinity. The concavity of the function allows only for one point for the derivative to vanish. Since this point happens to be the end of the interval, i.e., at~${r=1}$, so the function will not have a turning point in this range. This means that it starts off from~${1-L}$ at~${r=0}$ and reaches zero at~${r=1}$. So the function is nonpositive throughout this interval.
\end{enumerate}
So we have shown that for the case of~${q=1}$, the following holds: 
\eq{
\begin{cases}
r>1  \Longrightarrow m>0 \\
r=1  \Longrightarrow m=0 \\
r<1  \Longrightarrow m<0 \\
\end{cases}
}

\section{Discussion on the Solution of the Difference Equation and the Mean State for Open-ended Boundary Condition} \label{app:open_end}
Since the boundary condition is open-ended, at the top we have $s_0=1$ and at the bottom-most class, $s_L$ follows~\eqref{sdot4}, which simplifies to the following:
\al{
-\left(\frac{q \alpha + \beta}{q \alpha + \beta + p(y)} \right) s_L 
+  \left( \frac{\beta}{q \alpha + \beta + p(y)}\right) s_{L-1}=0
,}
since there is no bottom class for the $L$th class. This simplifies to: 
\eq{
r s_{L-1} - (1+r) s_L=0,}
which yields the following
\eq{
r \bigg[ K_1+K_2 r^{L-1} \bigg]  - (1+r) \bigg[ K_1+K_2 r^L \bigg] =0.
}
Replacing $K_1$ with $1-K_2$, we have
\eq{
\begin{cases}
K_1=  \displaystyle  \frac{-r^{L+1}}{1-r^{L+1}} \\ \\
K_2 =  \displaystyle  \frac{1}{1-r^{L+1}}.
\end{cases}
}
The steady state solution then becomes
\eq{
s_y=  \displaystyle  \frac{r^{L+1}-r^y}{r^{L+1}-1}
.}
Along the lines of the previous case, the mean state becomes
\eqq{
 \displaystyle m= \displaystyle \frac{r^{L+1}}{r^{L+1}-1} -  \displaystyle  \frac{\big[ (qr)^{L+1}-1\big] (q-1) }{(qr-1)(q^{L+1}-1)(r^{L+1}-1)}
.}{m_case2}
Note that the case of~${q=1}$ is singular and will be analyzed  afterwards. Taking the common denominator, we arrive at
\al{
m=\frac{r^{L+1} (qr-1)(q^{L+1}-1)- (q^{L+1}r^{L+1}-1)(q-1)}{(r^{L+1})(qr-1)(q^{L+1}-1)}
.}
The denominator has two poles, one at~${r=1}$ and the other at~${r=\frac{1}{q}}$. The sign of the denominator is negative in the interval between the poles, and is positive otherwise. So we have to show that the numerator has the same sign in these three ranges, so that~$m$ will be positive everywhere. Let us denote the denominator by~${h(\cdot)}$. Note that we have
\eq{
\begin{cases}
 \displaystyle h(1,q)=0 \\ \\
h \displaystyle \left( \frac{1}{q} ,q \right) = 0 \\ \\
 \displaystyle h(0,q)=q-1~>1.
\end{cases}
}
Taking the derivative, we have
\al{
\rond{h(r,q)}{r} &= (L+1)r^L(qr-1)(q^{L+1}-1)\nonumber \\
&+r^{L+1}q(q^{L+1}-1)-q^{L+1}(L+1)(q-1)r^L
.}
Evaluating this at~${r=1}$ yields
\al{
\left.  \rond{h(r,q)}{r} \right|_{r=1} &= (L+1)(q-1)(q^{L+1}-1) \nonumber \\
&+q(q^{L+1}-1)-(L+1)(q-1)q^{L+1} \nonumber \\
&= q^{L+2}-q(L+2)+(L+1)
.}
Note that this expression is zero at~${q=1}$, and its derivative with respect to~$q$ is
\al{
\left. \rondx{h(r,q)}{r}{q} \right|_{r=1} = (L+2) (q^{L+1}-1) >0
.}
So the derivative at~${r=1}$ is positive for all~${q>1}$. Now we evaluate the derivative at~${r=\frac{1}{q}}$. We obtain
\al{
\left.  \rond{h(r,q)}{r} \right|_{r=1/q} &\!\!\!\!\!\!= q^{-L}(q^{L+1}-1)-(L+1)(q^{L+2}-q^{L+1}) q^{-L} \nonumber \\
&\!\!\!\!\!\!\!\!\!\!\!\!\!\!\!= q^{-L} \bigg[  q^{L+1}-1-(L+1)q^{L+2}+(L+1)q^{L+1}            \bigg] \nonumber \\
&\!\!\!\!\!\!\!\!\!\!\!\!\!\!\!= -q^{-L}  \bigg[  (L+1)q^{L+2} -(L+2)q^{L+1}+1 \bigg] 
.}
We intend to show that this is negative, which is equivalent to the positivity of the term inside the brackets. Let us define
\eq{
\theta(q) \stackrel{\text{def}}{=}  (L+1)q^{L+2} -(L+2)q^{L+1}+1.
}
It is readily observable that~${\theta(1)=0}$. Also, taking the derivative, we obtain
\al{
\frac{d}{dq} \theta(q) &=  (L+2)(L+1)q^{L+1} -(L+2)(L+1)q^{L} \nonumber \\
&= (L+2)(L+1)q^L(q-1) >0.
}
Since the derivate is always positive for~${q>1}$, and the value of the function at~${q=1}$ is zero, so the function itself is positive in this range. 

So far we have seen that the derivative of~${h(r,q)}$ is negative at~${r=\frac{1}{q}}$, and is positive at~${r=1}$. The value of the function is zero at both points. If we show that the derivative only vanishes once in this interval, we can conclude that the function is negative between these two points, as desired, so that the mean state would be positive for all value of $r$. 

Setting the derivative equal to zero, we find
\al{
&\rond{h(r,q)}{r} = 0 \Longrightarrow (L+1)r^L(qr-1)(q^{L+1}-1)\nonumber \\
&+r^{L+1}q(q^{L+1}-1)-q^{L+1}(L+1)(q-1)r^L=0  
.}
Rearranging the terms and separating different powers of~$r$ gives
\al{
&r^{L+1} \bigg[ q(q^{L+1}-1)+q(L+1)(q^{L+1}-1) \bigg] \nonumber \\
&= r^L \bigg[ q^{L+1}(L+1)(q-1)+(L+1)  (q^{L+1}-1)\bigg] 
,}
which yields the only root
\al{
r= \frac{q^{L+1}(L+1)(q-1)+(L+1)  (q^{L+1}-1)}{q(q^{L+1}-1)+q(L+1)(q^{L+1}-1)}
.}
This can be simplified to 
\al{
r= \frac{(L+1)  (q^{L+2}-1)}  {q(L+2)(q^{L+1}-1)}
.}
This is the only root in the derivative. Now let us show that it is between the two roots, that is, 
\al{
\frac{1}{q} < \frac{(L+1)  (q^{L+2}-1)}  {q(L+2)(q^{L+1}-1)} <1
.}
This is equivalent to the following system of inequalities
\eq{
\begin{cases}
q^{L+2}-q(L+2)+(L+1)>0 \\
(L+1)q^{L+2}-(L+2)q^{L+1}+1>0.
\end{cases}
}
The second one is~$\theta(r)$, the positivity of which was proven above. For the first one, first note that its value is zero at~$q=1$. Its derivative is
\eq{
(L+2)q^{L+1}-(L+2)= (L+2) (q^{+1}-1) >0.
}
So we have found that the numerator of~\eqref{m_case2} has a single turning point between~${r=\frac{1}{q}}$ and~${r=1}$, and it is negative at the former and positive at the latter. Hence the numerator is negative between these points and positive otherwise, same as the denominator. So we conclude that the mean state is always nonnegative, regardless of~$r$. 

Now let us examine the peculiar case of~$q=1$. In this case, the mean state is
\al{
 \displaystyle m&= \displaystyle \frac{r^{L+1}}{r^{L+1}-1} -  \displaystyle  \frac{\big[ (r)^{L+1}-1\big]  }{(r-1)(r^{L+1}-1)(L+1)}  \nonumber \\
&=\displaystyle \frac{r^{L+1}}{r^{L+1}-1}-  \displaystyle  \frac{1  }{(r-1)(L+1)}
\nonumber \\
&= \frac{r^{L+2}(L+1)-r^{L+1}(L+2)+1}   {(r^{L+1}-1)(r-1)(L+1)}
.}

 The denominator is always positive, since the two factors flip sign at the same point. We require to verify that the numerator is always nonnegative. Let us define
\eq{
\eta(r) \stackrel{\text{def}}{=} r^{L+2}(L+1)-r^{L+1}(L+2)+1
.}
The derivative is
\eq{
\frac{d}{dr} \eta(r)=(L+2)(L+1)r^L(r-1)
.}
Since the function is zero at~${r=1}$ and the derivative is always positive for~${r>1}$, so the mean state is positive for this range. 

Now to find the sign of the mean state for~${m<1}$, note that the derivative is negative in this range, and there is only two turning points, which are~${r=0,1}$. the value of the mean state is positive for~${r=0}$, so the mean state is positive for~${r<1}$. 

We conclude that for the case of~${q=1}$, the mean state is positive regardless of the value of~$r$.


\begin{thebibliography}{2}%
\makeatletter
\providecommand \@ifxundefined [1]{%
 \@ifx{#1\undefined}
}%
\providecommand \@ifnum [1]{%
 \ifnum #1\expandafter \@firstoftwo
 \else \expandafter \@secondoftwo
 \fi
}%
\providecommand \@ifx [1]{%
 \ifx #1\expandafter \@firstoftwo
 \else \expandafter \@secondoftwo
 \fi
}%
\providecommand \natexlab [1]{#1}%
\providecommand \enquote  [1]{``#1''}%
\providecommand \bibnamefont  [1]{#1}%
\providecommand \bibfnamefont [1]{#1}%
\providecommand \citenamefont [1]{#1}%
\providecommand \href@noop [0]{\@secondoftwo}%
\providecommand \href [0]{\begingroup \@sanitize@url \@href}%
\providecommand \@href[1]{\@@startlink{#1}\@@href}%
\providecommand \@@href[1]{\endgroup#1\@@endlink}%
\providecommand \@sanitize@url [0]{\catcode `\\12\catcode `\$12\catcode
  `\&12\catcode `\#12\catcode `\^12\catcode `\_12\catcode `\%12\relax}%
\providecommand \@@startlink[1]{}%
\providecommand \@@endlink[0]{}%
\providecommand \url  [0]{\begingroup\@sanitize@url \@url }%
\providecommand \@url [1]{\endgroup\@href {#1}{\urlprefix }}%
\providecommand \urlprefix  [0]{URL }%
\providecommand \Eprint [0]{\href }%
\providecommand \doibase [0]{http://dx.doi.org/}%
\providecommand \selectlanguage [0]{\@gobble}%
\providecommand \bibinfo  [0]{\@secondoftwo}%
\providecommand \bibfield  [0]{\@secondoftwo}%
\providecommand \translation [1]{[#1]}%
\providecommand \BibitemOpen [0]{}%
\providecommand \bibitemStop [0]{}%
\providecommand \bibitemNoStop [0]{.\EOS\space}%
\providecommand \EOS [0]{\spacefactor3000\relax}%
\providecommand \BibitemShut  [1]{\csname bibitem#1\endcsname}%
\let\auto@bib@innerbib\@empty
\bibitem [{Note1()}]{Note1}%
  \BibitemOpen
  \bibinfo {note} {In this paper, it is tacitly assumed that there is a unique
  classifier of the members of society, based on which the levels are assigned.
  So the analysis is more applicable to situations where there is a dominant
  classification factor. For example, in an ideal-typical chain of command,
  one's rank is the determinant of influence, regardless of other
  socio-economic factors.}\BibitemShut {Stop}%
\bibitem [{Note2()}]{Note2}%
  \BibitemOpen
  \bibinfo {note} {Let us mention that Weber's social classification adds other
  factors to that of Marx, such as status groups, yielding a classification
  with at least two dimensions, altering the consequences~\cite {weber7,
  giddens}.}\BibitemShut {Stop}%
\end{thebibliography}%


\begin{thebibliography}{99}


\bibitem{liggett}
T. M. Liggett,  \emph{Interacting Particle Systems}, Springer, New York (1985). 


%
%
%
%
%
%
%
%


\bibitem{new1}
M. F.  Laguna , G. Abramson,  J. R.  Iglesias, Eur. Phys. J. B (2013) 86: 202. 

\bibitem{galam}
S. Galam, Y. Gefen, Y. Shapir, J. Math. Soci.,  \textbf{ 9},    1 (1982),  1-13. 


\bibitem{lewenstein}
M. Lewenstein, A. Nowak, B. Latane, Phys. Rev. A,  \textbf{ 45} (1992),  763-776. 

\bibitem{sznajd}
K. Sznajd-Weron, J. Sznajd,  Int. J  Mod. Phys. C.,  \textbf{ 11},  6 (2000), 1157-1165.
\bibitem{sznajd2}
K. Sznajd-Weron, ``Sznajd model and its applications.,  	Acta Physica Polonica B, vol.36, no. 8 (2005).

\bibitem{bernardes}
A.T. Bernardes, D. Stauffer and J. Kertesz,  Europ. Phys. J.  B, 
 \textbf{ 25},  1 (2002),  123-127.

\bibitem{contucci}
P. Contucci, S. Ghirlanda,  Quality  and Quantity,  \textbf{  41},   4 (2007),  569-578.

\bibitem{bornoldt}
S. Bornholdt, 
Int.   J. of Modern Physics C.,  \textbf{ 12},  5 (2001),  667-674. 

\bibitem{galam2}
R. Florian, S. Galam,  Europ. Phys. J.  B,  \textbf{ 16},  1 (2000),  189-194.


\bibitem{stauffer}
G. Zaklan, F. Westerhoff, D. Stauffer,  J.  of Econ. Interaction and Coordination,  \textbf{ 4},  1 (2009),  1-14. 


\bibitem{barra}
A. Barra, E. Agliari,  Physica A,  \textbf{ 391},    10 (2012),  3017-3026. 




\bibitem{schweitzer2}
F. Schweitzer,\emph{Modelling Complexity in Economic and Social Systems}, World Scientific, Singapore, 2002. 

\bibitem{krapivsky}
P. L. Krapivsky, S. Redner,  Phys. Rev. Lett.,  \textbf{ 90},    23  (2003).

\bibitem{vazquez2}
F. Vazquez, P. L. Krapivsky, S. Redner,   J.  of Phys.  A,  \textbf{ 36}, 3 (2003)

\bibitem{slanina}
F. Slanina, H. Lavicka,  Europ. Phys. J.  B,  \textbf{ 35},   2 (2003),  279-288.




\bibitem{martins}
A. C .R. Martins,   Phys. Rev. E,  \textbf{ 78},    3(2008).

\bibitem{delre}
S. A. Delre, W. Jager, T. Bijmolt, M. Janssen,   J.  of Product Innov. Manag.,  \textbf{ 27},    2  (2010),  267-282. 



\bibitem{naim1}
E. Ben-Naim,  EPL,  \textbf{ 69}, 5 (2005)

\bibitem{masuda}
N. Masuda, N. Gibert, S. Redner,   Phys. Rev. E, \textbf{82},    1 (2010)

\bibitem{lambiotte}
R. Lambiotte, S. Render, EPL,  \textbf{ 82},  1 (2008)
%

\bibitem{vilone}
C. Castellano, D. Vilone, A. Vespignani,  EPL,  \textbf{ 63},  1 (2003).



\bibitem{galam_rev}
S. Galam, Int. Jl of Modern Phys. C, Vol. 19, no. 03 (2008): 409-440.
\bibitem{castel_rev}
C. Castellano, S.  Fortunato, and V. Loreto, Rev. of mod. phys., vol.  81, no. 2 (2009): 591.


\bibitem{galambook}
S. Galam,  \emph{Sociophysics: A Physicist's Modelling of Psycho-political Phenomena}, Springer (2012).




\bibitem{Zgalam1}
S. Galam, J.  Math.  Psych., \textbf{30}, 4 (1986), 426-434. 
\bibitem{Zgalam2}
S. Galam, J. Soci. Complexity, \textbf{2}, 2 (2006), 62-75. 
\bibitem{Zgalam3}
S. Galam, J. Stat. Phys., \textbf{61}, 1990. 
\bibitem{Zgalam4}
S. Galam, Int. J. Of General Sys., Vol. 18, no. 3 (1991), 191-200.








\bibitem{sherif}
M. Sherif, \emph{The Psychology of Social Norms}, Oxford, England: Harper (1936). 
\bibitem{asch}
S. E. Asch, In ``Groups, Leadership and Men'', H. Guetzkow (ed.),  Carnegie Mellon Univ. Press (1951),  177-190.
\bibitem{milgram}
S. Milgram,   J.  of Abnormal and Soci.Psych.,  \textbf{ 64},    4 (1963),  371-378.
\bibitem{zimbardo}
C. Haney, C. Banks, P. Zimbardo, Int.  J.  of Criminology and Penology,  \textbf{ 1} (1973),  69-97.
\bibitem{kelman}
H. C. Kelman, H. V. Lee, \emph{Crimes of Obedience: Toward a Social Psychology of Authority and Responsibility}, Yale University Press, New Haven, CT, US (1989).
\bibitem{cialdini}
R. B. Cialdini, N. J. Goldstein, \emph{Social Influence: Compliance and Conformity}, Annu. Rev. of Psych.,  \textbf{ 55} (2004),  591-621. 
\bibitem{steve}
R. Brym,  J. Lie,  S.  Rytina. \emph{Networks, Groups, Bureaucracies, and Societies.} New Society, online chapter. Toronto: Nelson 5 (2008).










\bibitem{sood}
V. Sood, S. Redner,  Phys. Rev. Lett.,  \textbf{ 94},    17  (2005). 
\bibitem{vazquez}
F. Vazquez, S. Redner, J.  of Phys. A,  \textbf{ 37},  35 (2004). 
\bibitem{volovik}
D. Volovik, M. Mobilia, S. Redner,  Europhys. Lett.,  \textbf{ 85},  4 (2009). 
\bibitem{castellano}
C. Castellano, M. A Munoz, R. Pastor-Satorras, Phys. Rev. E,  \textbf{ 80},    4 (2009).
\bibitem{weighted1}
A. Baronchelli, C. Castellano, R. Pastor-Satorras,  Phys. Rev. E \textbf{83}, 6 (2011),  066117.
%




\bibitem{caccioli}
F. Caccioli,  L. Dall'Asta, T. Galla, T. Rogers,  Phys. Rev E 87.5 (2013): 052114.


\bibitem{contra}
 S. Do Yi,  S. K. Baek,  C. P. Zhu, B. J. Kim,  Phys. Rev E 87.1 (2013):  012806.



\bibitem{castel_gen}
C. Castellano,,  R. Pastor-Satorras,  Phys. Rev E 86.5 (2012): 051123.


\bibitem{rec1} 
M. Balb{\'a}s Gambra, E. Frey, Eur. Phys. J. B 83, 507–518 (2011). 
 
\bibitem{rec2}
P.  Nyczka, K. Sznajd-Weron, J.  Cisło, Phys. Rev E 86.1 (2012):  011105.

\bibitem{rec3}
G. A. B{\"o}hme, T. Gross, Rev E 85.6 (2012):  066117.

\bibitem{rec4}
C. J. Tessone, R. Toral, Eur. Phys. J. B 71, 549–555 (2009).


\bibitem{rec5}
S. M. Krause, S. Bornholdt, Phys. Rev E 86.5 (2012):   056106 .


\bibitem{rec6}
C. Biely, R. Hanel, S. Thurner, Eur. Phys. J. B 67, 285–289 (2009). 















\bibitem{weber1}
M. Weber, \emph{ The theory of social and economic organization}. Vol. 93493. Free Press (1997).

\bibitem{merton}
R. K. Merton, R. K. Gray,  B. Hockey, H. C. Selvin, \emph{Reader in bureaucracy},  Glencoe, Ill.: Free Press, 1952.

\bibitem{blau1}
P. M. Blau, \emph{Bureaucracy in modern society.} Crown Publishing Group, Random House (1956).

\bibitem{blau2}
P. M. Blau,  Am. J of Soci. (1968), 453-467.

\bibitem{peter}
L. J. Peter, R. Hull, L. Frey, \emph{The peter principle}, W. Morrow (1969).


\bibitem{ouchi}
W. G. Ouchi, Acad. of Manag. J., \textbf{21}, 2(1978), Acad. of Manag.

\bibitem{michels}
R. Michels, \emph{ Political parties: A sociological study of the oligarchical tendencies of modern democracy}. Hearst's International Library Company (1915).



\bibitem{greshon}
R. A. Gershon, \emph{The transnational media corporation: Global messages and free market competition},  Lawrence Erlbaum (1996).
\bibitem{shah}
A. Shah, \emph{Media conglomerates, mergers, concentration of ownership}, Global Issues (2009).

\bibitem{noam}
E. M. Noam, \emph{Media ownership and concentration in America},  Oxford University Press, USA (2009).


\bibitem{chomsky}
E. S. Herman, and N. Chomsky. \emph{Manufacturing consent: The political economy of the mass media.},  Pantheon (2002).
\bibitem{bagdakian}
B. H. Bagdikian.\emph{ The new media monopoly}. Beacon Press (2004)
\bibitem{anderson}
R. Andersen  and L. Strate, eds. \emph{Critical studies in media commercialism.} Oxford University Press (2000).


%

\bibitem{calhoun}
C. Calhoun, J. Gerteis, J. Moody, S. Pfaff, K. Schmidt, I. Virk, \emph{Classical Sociological Theory}, Blackwell Publishing, Malden, MA (2002). 
\bibitem{appelrouth}
S. Appelrouth, L. D. Edles, \emph{Classsical and Contemporary Sociological Theory: Text and Readings}, Los Angeles, CA (2008).
\bibitem{coser}
L. A. Coser, \emph{Masters of sociological thought: Ideas in historical and social context},  Harcourt Brace Jovanovich, New York, 1971

%
%
%

\bibitem{mills1}
C. Wright Mills, \emph{The power elite}, Oxford University Press, USA (2000).

\bibitem{dye}
T. R. Dye, \emph{Top down policymaking},  Chatham House Pub (2001).

\bibitem{elmer}
E. E. Schattschneider,  \emph{The Semi-Sovereign People: A Realist's View of Democracy in America.}, Wadsworth Publishing (1975).



\bibitem{weber7}
M. Weber, ``Max Weber: selections in translation", Cambridge University Press (1978).

\bibitem{giddens}
A. Giddens,  ``Marx, Weber, and the development of capitalism.", Sociology, Vol.  4, no. 3 (1970): 289-310.




%
%
%
%


%
%
%
%
%
%
%






%
%
%
%
%
%
%

%



%
%







%
%
%
%






\bibitem{polyanin}
A. D. Polyanin, \emph{Handbook of linear partial differential equations
for engineers and scientists}, Boca Raton: Chapman and Hall/CRC (2002).


\bibitem{bender}
C. M. Bender, S. A. Orszag,\emph{ Advanced Mathematical
Methods for Scientists and Engineers} , McGraw-Hill, New
York (1978).
















\end{thebibliography}
\end{document}